\documentclass[twocolumn,showpacs,preprintnumbers,amssymb,nofootinbib]{revtex4}

\usepackage{graphicx}
\usepackage{bm} 
\begin{document}

\title{The extremal limits of the C-metric:
Nariai, Bertotti-Robinson and anti-Nariai C-metrics}

\author{\'Oscar J. C. Dias}
\email{oscar@fisica.ist.utl.pt}
\affiliation{ Centro
Multidisciplinar de Astrof\'{\i}sica - CENTRA, \\Departamento de
F\'{\i}sica, Instituto Superior T\'ecnico, \\Av. Rovisco Pais 1,
1049-001 Lisbon
}
\author{Jos\'e P. S. Lemos}
\email{lemos@physics.columbia.edu} \affiliation{
Department of Physics, Columbia University, New York, NY 10027
\\ \&\\
Centro
Multidisciplinar de Astrof\'{\i}sica - CENTRA, Departamento de
F\'{\i}sica, Instituto Superior T\'ecnico, Av. Rovisco Pais 1,
1049-001 Lisbon,
}

\date{\today}
\begin{abstract}
In two previous papers we have analyzed the C-metric in a
background with a cosmological constant $\Lambda$, namely the de
Sitter (dS) C-metric ($\Lambda>0$), and the anti-de Sitter (AdS)
C-metric ($\Lambda<0$), extending thus the original work of
Kinnersley and Walker for the C-metric in flat spacetime
($\Lambda=0$). These exact solutions describe a pair of
accelerated black holes in the flat or cosmological constant
background, with the acceleration $A$ being provided by a strut
in-between that pushes away the two black holes or, alternatively,
by strings hanging from infinity that pull them in. In this paper
we analyze the extremal limits of the C-metric in a background
with generic cosmological constant $\Lambda>0,\,\Lambda=0,\,{\rm
and}\;\Lambda<0$.  We follow a procedure first introduced by
Ginsparg and Perry in which the Nariai solution, a spacetime which
is the direct topological product of the 2-dimensional dS and a
2-sphere, is generated from the four-dimensional dS-Schwarzschild
solution by taking an appropriate limit, where the black hole
event horizon approaches the cosmological horizon.  Similarly, one
can generate the Bertotti-Robinson metric from the
Reissner-Nordstr\"{o}m  metric by taking the limit of the Cauchy
horizon going into the event horizon of the black hole, as well as
the anti-Nariai by taking an appropriate solution and limit. Using
these methods we generate the C-metric counterparts of the Nariai,
Bertotti-Robinson and anti-Nariai solutions, among others. These
C-metric counterparts are conformal to the product of two
2-dimensional manifolds of constant curvature, the conformal
factor depending on the angular coordinate. In addition, the
C-metric extremal solutions have a conical singularity at least at
one of the poles of their angular surfaces.  We give a physical
interpretation to these solutions, e.g., in the Nariai C-metric
(with topology $dS_2\times \tilde{S}^2$) to each point in the
deformed 2-sphere $\tilde{S}^2$ corresponds a $dS_2$ spacetime,
except for one point which corresponds a $dS_2$ spacetime with an
infinite straight strut or string. There are other important new
features that appear. One expects that the solutions found in this
paper are unstable and decay into a slightly non-extreme black
hole pair accelerated by a strut or by strings. Moreover, the
Euclidean version of these solutions mediate the quantum process
of black hole pair creation, that accompanies the decay of the dS
and AdS spaces.
\end{abstract}


\maketitle

\section{\label{sec:Int}Introduction}

Three important exact solutions of general relativity are de
Sitter (dS) spacetime which is a spacetime with positive
cosmological constant ($\Lambda>0$), Minkowski (or flat) spacetime
with $\Lambda=0$, and anti-de Sitter (AdS) spacetime with negative
cosmological constant ($\Lambda<0$). These stainless spacetimes,
with trivial topology $\mathbb{R}^4$, have nonetheless a rich
internal structure best displayed through the Carter-Penrose
diagrams \cite{hawkingellis}.  They also serve as the background
to spacetimes containing black holes which are then asymptotically
dS, flat, or AdS. These black holes in background spacetimes with
a cosmological constant - Schwarzschild, Reissner-Nordstr\"om,
Kerr, and Kerr-Newman -  have a complex causal and topological
structure well described in \cite{carter}.

There are other very interesting solutions of general relativity
with generic cosmological constant, that are neither pure nor
contain a black hole, and somehow can be considered intermediate
type solutions. These are the Nariai, Bertotti-Robinson, and
anti-Nariai solutions \cite{KramSMH}.  The Nariai solution
\cite{Nariai,Nariai2} solves exactly the Einstein equations with
$\Lambda>0$, without or with a Maxwell field, and has been
discovered by Nariai in 1951 \cite{Nariai}. It is the direct
topological product of $dS_2 \times S^2$, i.e., of a
(1+1)-dimensional dS spacetime with a round 2-sphere of fixed
radius. The Bertotti-Robinson solution \cite{BertRob} is an exact
solution of the Einstein-Maxwell equations with any $\Lambda$, and
was found independently by Bertotti and Robinson in 1959. It is
the direct topological product of $AdS_2 \times S^2$, i.e., of a
(1+1)-dimensional AdS spacetime with a round 2-sphere of fixed
radius. The anti-Nariai solution, i.e., the AdS counterpart of the
Nariai solution, also exists \cite{CaldVanZer} and is an exact
solution of the Einstein equations with $\Lambda<0$, without or
with a Maxwell field. It is the direct topological product of
$AdS_2 \times H_2$, with $H_2$ being a 2-hyperboloid.

Three decades after Nariai's paper, Ginsparg and Perry
\cite{GinsPerry} connected the Nariai solution with the
Schwarzschild-dS solution. They showed that the Nariai solution
can be generated from a near-extreme dS black hole, through an
appropriate limiting procedure in which the black hole horizon
approaches the cosmological horizon. A similar extremal limit
generates the Bertotti-Robinson solution and the anti-Nariai
solution from an appropriate near extreme black hole (see, e.g.
\cite{CaldVanZer}). One of the aims of Ginsparg and Perry was to
study the quantum stability of the Nariai and the Schwarzschild-dS
solutions \cite{GinsPerry}.  It was shown that the Nariai solution
is in general unstable and, once created, decays through a quantum
tunnelling process into a slightly non-extreme black hole pair
(for a complete review and references on this subject see, e.g.,
Bousso \cite{Bousso60y}, and later discussions on our paper).  The
same kind of process happens for the Bertotti-Robinson and
anti-Nariai solutions.

There is yet another class of related metrics, the C-metric class,
which represent not one, but two black holes, being accelerate
apart from each other. These black holes can also inhabit a dS,
flat or AdS background.  Following the approach of Kinnersley and
Walker \cite{KW}
for the $\Lambda=0$ C-metric, we have analyzed in detail,
in two previous papers
\cite{OscLem_dS-C,OscLem_AdS-C}, the
physical interpretation and properties of the dS C-metric, i.e.,
the C-metric with $\Lambda>0$ \cite{OscLem_dS-C}, and of the AdS
C-metric, i.e., the C-metric with $\Lambda<0$ \cite{OscLem_AdS-C}.
As occurs with the flat C-metric,
the cosmological  constant C-metric describes a pair of accelerated black
holes with the acceleration that drives them away being provided
by a strut or by strings (although in the AdS C-metric case, this
is true only when the acceleration $A$ and the cosmological
constant are related by $A>|\Lambda|/3$ \cite{OscLem_AdS-C}). When
the acceleration is zero the C-metric reduces to a single
non-accelerated black hole with the usual properties.

It is therefore of great interest to apply the Ginsparg-Perry
procedure to these metrics in order to find a new set of exact
solutions with a clear physical and geometrical interpretation.  In
this paper we address this issue of the extremal limits of the
C-metric with a generic $\Lambda$ following \cite{GinsPerry}, in order
to generate the C-metric counterparts ($A \neq 0$) of the Nariai,
Bertotti-Robinson and anti-Nariai solutions ($A=0$),
among others.

The plan of this paper is as follows. In section
\ref{sec:Nariai,BR}, we describe the main features of the Nariai,
Bertotti-Robinson and anti-Nariai solutions. In section
\ref{sec:ExtLim dS}, we analyze the extremal limits of the dS
C-metric. We generate the Nariai C-metric, the Bertotti-Robinson
dS C-metric, and the Nariai Bertotti-Robinson dS C-metric. We
study the topology and the causal structure of these solutions,
and we give a physical interpretation. In section \ref{sec:ExtLim
L=0} we present the extremal limits of the flat C-metric and Ernst
solution, which are obtained from the solutions of section
\ref{sec:ExtLim dS} by taking the direct $\Lambda=0$ limit. The
Euclidean version of one of these $\Lambda=0$ solutions has
already been used previously, but we discuss two new solutions
that have not been discussed previously. In section
\ref{sec:ExtLim AdS}, we discuss the extremal limits of the AdS
C-metric and, in particular, we generate the anti-Nariai C-metric.
Finally, in section \ref{sec:Conc} concluding remarks are
presented.

\section{\label{sec:Nariai,BR}Extremal limits of black hole solutions in
\lowercase{d}S, flat, and A\lowercase{d}S
spacetimes: The Nariai, Bertotti-Robinson and anti-Nariai solutions}
In this section we will describe the main features of the Nariai,
Bertotti-Robinson and anti-Nariai solutions. The extremal limits
of the C-metric that will be generated in later sections reduce to
these solutions when the acceleration parameter $A$ is set to
zero.

\subsection{\label{sec:Nariai}The Nariai solution}

The neutral Nariai solution has been first introduced
by Nariai \cite{Nariai,Nariai2}. It
satisfies the Einstein field
equations  in
a positive cosmological constant $\Lambda$
background, $G_{\mu\nu}+\Lambda
g_{\mu\nu}=0$, and is given by
\begin{equation}
d s^2  = \Lambda^{-1}(- \sin^2\chi\, d\tau^2 +d\chi^2
+d \theta^2+\sin^2\theta\,d\phi^2)\,,\label{mNariai}
\end{equation}
where $\chi$ and
$\theta$ both run from $0$ to $\pi$, and $\phi$ has period $2\pi$.

The electromagnetic extension of the Nariai solution has been
introduced by Bertotti and Robinson \cite{BertRob}.
Its gravitational field is given by
\begin{equation}
d s^2 =  \frac{{\cal R}_0^{\:2}}{ {\cal K}_0}\left (-
\sin^2\chi\, d\tau^2 +d\chi^2 \right )
+{\cal R}_0^{\:2} \left ( d \theta^2+\sin^2\theta\,d\phi^2 \right )\,
\label{qNariai}
\end{equation}
where ${\cal R}_0$ is a positive constant and constant
${\cal K}_0$ satisfies $0<{\cal K}_0\leq 1$, while the electromagnetic
field of the Nariai solution is
\begin{eqnarray}
F=q \sin \theta \, d\theta \wedge d\phi\,
 \label{MpotNariai}
\end{eqnarray}
in the purely magnetic case, and
\begin{eqnarray}
 F=\frac{q}{{\cal K}_0}\,\sin \chi \,d\tau \wedge d\chi \,
 \label{EpotNariai}
\end{eqnarray}
in the purely electric case, with $q$ being the electric or
magnetic charge, respectively. The cosmological constant and the
charge of the Nariai solution are related to ${\cal R}_0$ and
 ${\cal K}_0$ by
\begin{eqnarray}
& & \Lambda=\frac{1+{\cal K}_0}{2{\cal R}_0^{\:2}} \,, \nonumber \\
& & q^2=\frac{1-{\cal K}_0}{2}\,{\cal R}_0^{\:2}\,.
\label{relations}
\end{eqnarray}
Note that $0<{\cal K}_0\leq 1$, otherwise the charge is a complex
number. The charged Nariai solution satisfies the field equations
of the Einstein-Maxwell action in a positive cosmological constant
background, $G_{\mu\nu}+\Lambda g_{\mu\nu}=8\pi T_{\mu\nu}$, with
$T_{\mu\nu}$ being the energy-momentum tensor of the Maxwell
field. The neutral Nariai solution (\ref{mNariai}) is obtained
from the charged solution (\ref{qNariai})-(\ref{MpotNariai}) when
one sets ${\cal K}_0=1$. The $\Lambda=0$ limit of the Nariai
solution, which is Minkowski spacetime, is taken in the Appendix
\ref{sec:lim L=0 Nariai}. Through a redefinition of coordinates,
$\sin^2\chi=1-\frac{{\cal K}_0}{{\cal R}_0^{\:2}}\: R^2$ and
$\tau=\sqrt{\frac{{\cal K}_0}{{\cal R}_0^{\:2}}} \, T$, the
spacetime (\ref{qNariai}) can be rewritten in new static
coordinates as
\begin{eqnarray}
d s^2 = - N(R)\, dT^2 +\frac{dR^2}{N(R)}
+{\cal R}_0^{\:2}\left (d \theta^2+\sin^2\theta\,d\phi^2 \right ),
\label{qNariai2}
\end{eqnarray}
with
\begin{eqnarray}
N(R)=1-\frac{{\cal K}_0}{{\cal R}_0^{\:2}} \:R^2 \, ,
 \label{N}
\end{eqnarray}
and the electromagnetic field changes also accordingly to the
coordinate transformation. Written in these coordinates, we
clearly see that the Nariai solution is the direct topological
product of $dS_2 \times S^2$, i.e., of a (1+1)-dimensional dS
spacetime with a round 2-sphere of fixed radius ${\cal R}_0$. This
spacetime is homogeneous with the same causal structure as
(1+1)-dimensional  dS spacetime, but it is not an asymptotically
4-dimensional dS spacetime since the radius of the 2-spheres is
constant (${\cal R}_0$), contrarily to what happens in the dS
solution where this radius increases as one approaches infinity.

Another way \cite{Nariai2,Ortag} to see clearly the topological
structure of the Nariai solution is achieved by defining it
through its embedding in the flat manifold ${\mathbb{M}}^{1,5}$,
with  metric
\begin{eqnarray}
d s^2 = -dz_0^{\:\:2}+dz_1^{\:\:2}+dz_2^{\:\:2}+
dz_3^{\:\:2}+dz_4^{\:\:2}+dz_5^{\:\:2} \,.
\label{M1+5}
\end{eqnarray}
The Nariai 4-submanifold is determined by the two constraints
\begin{eqnarray}
& & -z_0^{\:\:2}+z_1^{\:\:2}+z_2^{\:\:2}= \ell^2 \:,
             \nonumber \\
& & z_3^{\:\:2}+z_4^{\:\:2}+z_5^{\:\:2}= {\cal R}_0^{\:2} \:,
\label{hypersurface}
\end{eqnarray}
where $\ell^2=  {\cal R}_0^{\:2}/{\cal K}_0$. The first of these
constraints defines the $dS_2$ hyperboloid and the second defines
the 2-sphere of radius ${\cal R}_0$. The parametrization of
${\mathbb{M}}^{1,5}$ given by $z_0=\sqrt{\ell^2-R^2}\sinh \left (
T/\ell \right )$, $z_1=\sqrt{\ell^2-R^2}\,\cosh \left ( T/\ell
\right)$, $z_2=R$, $z_3= {\cal R}_0 \sin\theta \cos\phi$, $z_4=
{\cal R}_0 \sin\theta \sin\phi$ and $z_5= {\cal R}_0 \cos\theta$,
induces the metric (\ref{qNariai2}) on the Nariai hypersurface
(\ref{hypersurface}).

Quite remarkably, Ginsparg and Perry \cite{GinsPerry} (see also
\cite{BoussoHawk}) have shown that the neutral Nariai solution
(\ref{mNariai}) can be obtained from the near-extreme
Schwarzschild-dS black hole through an appropriate limiting
procedure. By extreme we mean that the black hole horizon and the
cosmological horizon coincide. Hawking and Ross \cite{HawkRoss},
and Mann and Ross \cite{MannRoss} have concluded that a similar
limiting approach takes the near-extreme
dS$-$Reissner-Nordstr\"{o}m black hole into the charged Nariai
solution (\ref{qNariai}). In this case, by extreme we mean that
the cosmological and outer charged black hole horizons coincide.
We will make heavy use of this Ginsparg-Perry procedure later, so
we will not sketch it here. These relations between the
near-extreme dS black holes and the Nariai solutions are a priori
quite unexpected since (i) the dS black holes have a curvature
singularity while the Nariai solutions do not, (ii) the Nariai
spacetime is homogeneous unlike the dS black holes spacetimes,
(iii) the dS black holes approach asymptotically the 4-dimensional
dS spacetime while the Nariai solutions do not.
The Carter-Penrose diagram of the Nariai solution is equivalent to
the diagram of the (1+1)-dimensional dS solution, as will be
discussed in subsection \ref{sec:Gen Nariai}.

An important role played by the Nariai solution in physics is at
the quantum level (see Bousso \cite{Bousso60y} for a detailed
review of what follows). First, there is the issue of the
stability of the solution when perturbed quantically. Ginsparg and
Perry \cite{GinsPerry}, in the neutral case, and Bousso and
Hawking \cite{BoussoHawk} and Bousso \cite{BoussoDil}, in the
charged case, have shown that the Nariai solutions are quantum
mechanically unstable. Indeed, due to quantum fluctuations, the
radius ${\cal R}_0$ of the 2-spheres oscillates along the
non-compact spatial coordinate $\chi$ and the degenerate horizon
splits back into a black hole and a cosmological horizon. Those
2-spheres whose radius fluctuates into a radius smaller than
${\cal R}_0$ will collapse into the dS black hole interior, while
the 2-spheres that have a radius greater than ${\cal R}_0$ will
suffer an exponential expansion that generates an asymptotic dS
region. Therefore, the Nariai solutions are unstable and, once
created, they decay through the quantum tunnelling process into a
slightly non-extreme  dS$-$Reissner-Nordstr\"{o}m black hole pair.
This issue of the Nariai instability against perturbations and
associated evaporation process has been further analyzed by Bousso
and Hawking \cite{BoussoHawk-T}, by Nojiri and Odintsov
\cite{NojOd1}, and by Kofman, Sahni, and Starobinski
\cite{ShaKof}. Second, the Nariai Euclidean solution plays a
further role as an instanton, in the quantum decay of the dS
space. This decay of the dS space is accompanied by the creation
of a dS black hole pair, in a process that is the gravitational
analogue of the Schwinger pair production of charged particles in
an external electromagnetic field. Here, the energy necessary to
materialize the black hole pair and to accelerate the black holes
apart comes from the cosmological constant background. It is
important to note that not all of the dS black holes can be pair
produced through this quantum process of black hole pair creation.
Only those black holes that have regular Euclidean sections can be
pair created (the term regular is applied here in the context of
the analysis of the Hawking temperature of the horizons). The
Nariai instanton (regular Euclidean Nariai solution, that is
obtained from Eqs. (\ref{mNariai}) and (\ref{qNariai}) by setting
$\tau=i\bar{\tau}$), belongs to the very restrictive class of
solutions that are regular
\cite{MelMosRom,MannRoss,BoussoHawk,BooMann,VolkovWipf}, and can
therefore mediate the pair creation process in the dS background.
In the uncharge case the Nariai instanton is even the only
solution that can describe the pair creation of neutral black
holes. Another result at the quantum level by Medved \cite{Medv}
indicates that quantum back-reaction effects prevent a near
extreme dS black hole from ever reach a Nariai state of precise
extremality.

Other extensions of the Nariai solution are the dilaton charged
Nariai solution found by Bousso \cite{BoussoDil}, the rotating
Nariai solution studied by Mellor and Moss \cite{MelMosRom}, and
by Booth and Mann \cite{BooMann}, and solutions that describe
non-expanding impulsive waves propagating in the Nariai universe
studied by Ortaggio \cite{Ortag}.

\subsection{\label{sec:BR}The Bertotti-Robinson solution}

The simplest Bertotti-Robinson solution \cite{BertRob} (see also
\cite{CaldVanZer,Lap}) is a solution of the $\Lambda=0$
Einstein-Maxwell equations. Its gravitational field is given by
\begin{eqnarray}
& & d s^2  = \frac{1}{q^2}(- \sinh^2\chi\, d\tau^2 +d\chi^2)+
q^2(d
\theta^2+\sin^2\theta\,d\phi^2)\,, \nonumber \\
& &
 \label{br}
\end{eqnarray}
where $q$ is the charge of the solution, and $\chi$ is unbounded,
$\theta$ runs from $0$ to $\pi$ and $\phi$ has period $2\pi$. The
electromagnetic field of the Bertotti-Robinson solution is
\begin{eqnarray}
F=q \sin \theta \, d\theta \wedge d\phi \,,
\label{EpotBRmagnetic}
\end{eqnarray}
and
\begin{eqnarray}
 F=-q \,\sinh \chi \,d\tau \wedge d\chi \,,
  \label{EpotBRelectric}
\end{eqnarray}
in the magnetic and electric cases, respectively. Through a
redefinition of coordinates, $\sinh^2 \chi=R^2/q^2-1$ and
$\tau=T/q$, the spacetime (\ref{br}) can be rewritten in new
static coordinates as
\begin{eqnarray}
d s^2 = - N(R)\, dT^2 +\frac{dR^2}{N(R)}
 +q^2(d \theta^2+\sin^2\theta\,d\phi^2),
 \label{br2}
\end{eqnarray}
with
\begin{eqnarray}
N(R)=R^2/q^2-1 \, ,
 \label{N-br2}
\end{eqnarray}
and the electromagnetic field changes also accordingly to the
coordinate transformation. Written in these coordinates, we
clearly see that the Bertotti-Robinson solution is the direct
topological product of $AdS_2 \times S^2$, i.e., of a
(1+1)-dimensional AdS spacetime with a round 2-sphere of fixed
radius $q$.  Another way to see clearly the topological structure
of the Bertotti-Robinson solution is achieved by defining it
through its embedding in the flat manifold ${\mathbb{M}}^{2,4}$,
with metric
\begin{eqnarray}
 d s^2 = -dz_0^{\:\:2}+dz_1^{\:\:2}-dz_2^{\:\:2}+dz_3^{\:\:2}
 +dz_4^{\:\:2}+dz_5^{\:\:2} \,.
 \label{M1+5-br}
 \end{eqnarray}
The Bertotti-Robinson 4-submanifold is determined by the two
constraints
\begin{eqnarray}
& & -z_0^{\:\:2}+z_1^{\:\:2}-z_2^{\:\:2}= -q^2 \:,
             \nonumber \\
& & z_3^{\:\:2}+z_4^{\:\:2}+z_5^{\:\:2}=q^2 \,.
\label{hypersurface-br}
\end{eqnarray}
The first of these constraints defines the $AdS_2$ hyperboloid and
the second defines the 2-sphere of radius $q$. The parametrization
of ${\mathbb{M}}^{2,4}$,
 $z_0=\sqrt{R^2-q^2}\,\sinh \left ( T/q \right)$,
 $ z_1=\sqrt{R^2-q^2}\,\cosh \left ( T/q \right )$,
 $ z_2= R$,
 $ z_3= q \sin\theta \cos\phi$,
 $ z_4= q\sin\theta \sin\phi$ and
$ z_5= q \cos\theta$, induces the metric (\ref{br2}) on the
Bertotti-Robinson hypersurface (\ref{hypersurface-br}). Note that
since the parametrization
 $z_0=\sqrt{q^2+\tilde{R}^2}\,\sin(\tilde{T}/q)$,
 $z_1=\tilde{R}$ and
 $z_2=\sqrt{q^2+\tilde{R}^2}\,\cos(\tilde{T}/q)$,
also obeys (\ref{M1+5-br}) and the first constraint of
(\ref{hypersurface-br}), the Bertotti-Robinson solution
(\ref{br2}) can also be written as
 $d s^2 = - N(\tilde{R})\, d\tilde{T}^2
+d\tilde{R}^2 / N(\tilde{R})
 +q^2(d \theta^2+\sin^2\theta\,d\phi^2)\,,$ with
$N(\tilde{R})=\tilde{R}^2/q^2+1\,.$

Following a similar procedure to the one applied in the Nariai
solution, the Bertotti-Robinson solution can be obtained from the
near-extreme Reissner-Nordstr\"{o}m black hole through an
appropriate limiting procedure (here,  by extreme we mean that the
inner black hole horizon and the outer black hole horizon
coincide). Later on, we will make heavy use of the Ginsparg-Perry
procedure, so we will not sketch it here. Generalizations of the
Bertotti-Robinson solution to include a cosmological constant
background also exist \cite{BertRob}, and are an extremal limit of
the near-extreme Reissner-Nordstr\"{o}m black holes with a
cosmological constant. The Carter-Penrose diagram of the
Bertotti-Robinson solution (with or without $\Lambda$) is
equivalent to the diagram of the (1+1)-dimensional AdS solution,
as will be discussed in subsection \ref{sec:Gen BR dS}.

The Hawking effect in the Bertotti-Robinson universe has been
studied by Lapedes \cite{Lap}, and its thermodynamic properties
have been analyzed by Zaslavsky
\cite{Zasl}, and by Mann and Solodukhin \cite{MannSolod}. In
\cite{Navarro} the authors have shown that quantum back-reaction
effects prevent a near extreme charged black hole from ever reach
a Berttoti-Robinson state of precise extremality. Recently,
Ortaggio and Podolsk\'y \cite{OrtagPod} have found exact solutions
that describe non-expanding impulsive waves propagating in the
Bertotti-Robinson universe.

\subsection{\label{sec:anti-Nariai}The anti-Nariai solution}
The anti-Nariai solution has a gravitational field given by
\cite{CaldVanZer}
\begin{eqnarray}
ds^2  \!\!&=&\!\! \frac{{\cal R}_0^{\:2}}{{\cal K}_0} \left(
-\sinh^2\chi\, d\tau^2\!  + \! d\chi^2 \right )\!  + \! {\cal
R}_0^{\:2}\!  \left( d\theta^2\! +\! \sinh^2\theta\,d\phi^2
\right), \nonumber \\
& &
\label{qNariai-a}
\end{eqnarray}
where $\chi$ and $\theta$ are unbounded, $\phi$ has period
$2\pi$, ${\cal R}_0$ is a positive constant, and the constant
${\cal K}_0$ satisfies $1 \leq {\cal K}_0 <2$. The electromagnetic
field of the anti-Nariai solution is
\begin{eqnarray}
F=q \sinh \theta \, d\theta \wedge d\phi\,
 \label{MpotNariai-a}
\end{eqnarray}
in the purely magnetic case, and
\begin{eqnarray}
F=-\frac{q}{{\cal K}_0}\,\sinh \chi \, d\tau \wedge d\chi \,
\label{EpotNariai-a}
\end{eqnarray}
in the purely electric case, with $q$ being the magnetic or
electric charge, respectively. The cosmological constant,
$\Lambda<0$, and the charge of the anti-Nariai solution are
related to ${\cal R}_0$ and ${\cal K}_0$ by
\begin{eqnarray}
& & \Lambda=-\frac{1+{\cal K}_0}{2{\cal R}_0^{\:2}}<0
\,, \nonumber \\
& & q^2=\frac{{\cal K}_0-1}{2}\,{\cal R}_0^{\:2}\,.
\label{relations-a}
\end{eqnarray}
The neutral anti-Nariai solution ($q=0$) is obtained from the
charged solution (\ref{qNariai-a}) when one sets ${\cal K}_0=1$
which implies ${\cal R}_0^{\:2}=|\Lambda|^{-1}$.  The $\Lambda=0$
limit of the anti-Nariai solution, which is Minkowski spacetime,
is taken in the Appendix \ref{sec:lim L=0 Nariai}. The charged
anti-Nariai solution satisfies the field equations of the
Einstein-Maxwell action in a negative cosmological constant
background. Through a redefinition of coordinates, $\sinh^2\chi=1-
\frac{{\cal K}_0}{{\cal R}_0^{\:2}}\: R^2 $ and $\tau=
\sqrt{\frac{{\cal K}_0}{{\cal R}_0^{\:2}}} \: T$, the spacetime
(\ref{qNariai-a}) can be rewritten in static coordinates as
\begin{eqnarray}
d s^2 = - N(R)\, dT^2 +\frac{dR^2}{N(R)}
 +{\cal R}_0^{\:2}\left (d \theta^2+\sinh^2\theta\,d\phi^2 \right ),
  \nonumber \\
 \label{qNariai2-a}
\end{eqnarray}
with
\begin{eqnarray}
N(R)=-1+\frac{{\cal K}_0}{{\cal R}_0^{\:2}} \:R^2 \,,
 \label{N-a}
\end{eqnarray}
and the electromagnetic field changes also accordingly to the
coordinate transformation. Written in these coordinates, we
clearly see that the anti-Nariai solution is the direct
topological product of $AdS_2 \times H_2$, i.e., of a
(1+1)-dimensional AdS spacetime with a 2-hyperboloid of radius
${\cal R}_0$. It is a homogeneous spacetime with the same causal
structure as (1+1)-dimensional AdS spacetime, but it is not an
asymptotically 4-dimensional AdS spacetime since the size of the
2-hyperboloid is constant (${\cal R}_0$), contrarily to what
happens in the AdS solution where this radius increases as one
approaches infinity. Another way to see clearly the topological
structure of the anti-Nariai solution is achieved by defining it
through its embedding in the flat manifold ${\mathbb{M}}^{3,3}$
with metric
\begin{eqnarray}
d s^2 = -dz_0^{\:\:2}+dz_1^{\:\:2}-dz_2^{\:\:2}+dz_3^{\:\:2}
+dz_4^{\:\:2}-dz_5^{\:\:2} \,.
\label{M1+5-a}
\end{eqnarray}
The anti-Nariai 4-submanifold is determined by the two constraints
\begin{eqnarray}
& & -z_0^{\:\:2}+z_1^{\:\:2}-z_2^{\:\:2}= -\ell^2 \, , \nonumber \\
& & z_3^{\:\:2}+z_4^{\:\:2}-z_5^{\:\:2}= -{\cal R}_0^{\:2} \,.
\label{hypersurface-a}
\end{eqnarray}
where $\ell^2= {\cal R}_0^{\:2}/{\cal K}_0$. The first of these
constraints defines the $AdS_2$ hyperboloid and the second defines
the 2-hyperboloid of radius ${\cal R}_0$. The parametrization of
${\mathbb{M}}^{3,3}$, $z_0=\sqrt{R^2-\ell^2}\,\sinh \left ( T/\ell
\right)$, $z_1=\sqrt{R^2-\ell^2}\,\cosh \left ( T/\ell \right)$,
$z_2=R$, $z_3= {\cal R}_0 \sinh\theta \cos\theta$, $z_4= {\cal
R}_0 \sinh\theta \sin\theta$ and $z_5= {\cal R}_0 \cosh\theta$
induces the metric (\ref{qNariai2-a}) on the anti-Nariai
hypersurface (\ref{hypersurface-a}).

Having in mind the example of the Nariai solution, we may ask if
the anti-Nariai solution can be obtained, through a similar
limiting Ginsparg-Perry procedure, from a near-extreme AdS black
hole. There are black holes whose horizons have topologies
different from spherical, such as toroidal horizons \cite{toroidal},
and hyperbolical horizons \cite{topological}, also called
topological black holes. The AdS black hole that
generates the anti-Nariai solution is the hyperbolic one [as
is clear from the angular part of Eq. (\ref{qNariai2-a})],
and has a cosmological horizon. We will make a further reference to their
properties in section \ref{sec:ExtLim AdS}. The anti-Nariai
solution can be obtained from the extremal limit of these black
holes when the black hole horizon approaches the cosmological
horizon.

A further study of the anti-Nariai solution was done in
\cite{OrtagPod} where  non-expanding impulsive waves propagating
in the anti-Nariai universe are described.

\section{\label{sec:ExtLim dS}Extremal limits of the \lowercase{d}S C-metric}

In the last section we saw that there is an appropriate extremal
limiting procedure, introduced by Ginsparg and Perry
\cite{GinsPerry}, that generates from the near-extreme black hole
solutions the Nariai, Bertotti-Robinson and anti-Nariai solutions.
Analogously, we shall apply the procedure of \cite{GinsPerry} to
generate new exact solutions from the near-extreme cases of the dS
C-metric. Specifically the C-metric counterparts of the Nariai and
Bertotti-Robinson solutions are found using this method. When the
acceleration parameter $A$ is set to zero in these new solutions,
we will recover the Nariai and Bertotti-Robinson solutions. First
we will briefly review the dS C-metric \cite{OscLem_dS-C}.

\subsection{\label{sec:dS C-metric}The \lowercase{d}S C-metric}

\begin{figure} [t]
\includegraphics*[height=1.6in]{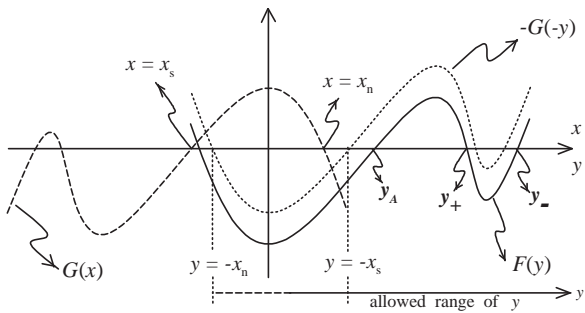}
\caption{\label{g3-Nariai} Shape of ${\cal G}(x)$ and ${\cal
F}(y)$ for a general non-extremal charged massive dS C-metric
studied in section \ref{sec:ExtLim dS}. The allowed range of $x$
is between $x_\mathrm{s}$ and $x_\mathrm{n}$ where ${\cal G}(x)$
is positive and compact. The permitted range of $y$ is $-x\leq y <
+\infty$. The presence of an accelerated horizon is indicated by
$y_A$ and the inner and outer charged horizons by $y_-$ and $y_+$.
In the Nariai case, $y_+$ and $y_A$ coincide (see section
\ref{sec:Gen Nariai}), in the Bertotti-Robinson case, $y_-$ and
$y_+$ coincide (see section \ref{sec:Gen BR dS}), and in the
Nariai Bertotti-Robinson case, $y_-$, $y_+$ and $y_A$ coincide
(see section \ref{sec:Gen Nariai-BR dS}). In the cases studied in
this paper, ${\cal G}(x)$ has only two real zeros, $x_\mathrm{s}$
and $x_\mathrm{n}$.
 }
\end{figure}

The massive charged dS C-metric has been found by Pleba\'nski and
Demia\'nski \cite{PlebDem}, and its gravitational field can be
written as (see, e.g., \cite{OscLem_dS-C})
\begin{equation}
 d s^2 = [A(x+y)]^{-2} (-{\cal F}dt^2+
 {\cal F}^{-1}dy^2+{\cal G}^{-1}dx^2+
 {\cal G}dz^2)\:,
 \label{C-metric}
 \end{equation}
 where
 \begin{eqnarray}
 & &{\cal F}(y) = -\frac{\Lambda+3A^2}{3A^2}
                     +y^2-2mAy^3+q^2A^2y^4, \nonumber \\
 & &{\cal G}(x) = 1-x^2-2mAx^3-q^2 A^2 x^4\:,
 \label{FG}
 \end{eqnarray}
 and the Maxwell field
 in the magnetic case is given by
\begin{eqnarray}
 F=-q\, dx\wedge d\phi \:,
\label{F-mag}
\end{eqnarray}
while in the electric case it is given by
 \begin{eqnarray}
 F=-q\, dt\wedge dy \:.
\label{F-el-Lorentz}
\end{eqnarray}
 This solution depends on four parameters namely, the cosmological
 constant $\Lambda$, $A>0$ which
 is the acceleration of the black holes, and $m$ and $q$ which are
 interpreted  as the ADM mass and electromagnetic charge of the
 non-accelerated black hole.  The general shape of ${\cal G}(x)$ and ${\cal
 F}(y)$ is represented in Fig. \ref{g3-Nariai}. The physical
 properties and interpretation of this solution have been analyzed
 by Dias and Lemos \cite{OscLem_dS-C}, and
 by  Podolsk\'y and Griffiths \cite{PodGrif2}.

The solution has a curvature singularity at $y=+\infty$, and $y$
must belong to the range $-x\leq y <+\infty$. The point $y=-x$
corresponds to a point that is infinitely far away from the
curvature singularity, thus as $y$ increases we approach the
curvature singularity and $y+x$ is the inverse of a radial
coordinate. At most, ${\cal F}(y)$ can have four real zeros which
we label in ascending order by $y_{\rm neg}<0<y_A\leq y_+ \leq
y_-$. The roots $y_-$ and $y_+$ are respectively the inner and
outer charged black hole horizons, and $y_A$ is an acceleration
horizon which coincides with the cosmological horizon and has a
non-spherical shape, although the topology is spherical.  The
negative root $y_{\rm neg}$ satisfies $y_{\rm neg}<-x$ and has no
physical significance, i.e., it does not belong to the range
accessible to $y$.

We will demand that $x$ belongs to the interval
$x_\mathrm{s}\leq x\leq x_\mathrm{n}$,
sketched in Fig. \ref{g3-Nariai}
where ${\cal G}(x)\geq 0$. By doing this we guarantee that the
metric has the correct signature $(-+++)$ [see Eq.
(\ref{C-metric})] and that the angular surfaces $t=$const and
$y=$const are compact. For a reason that we shall explain soon we
will label these surfaces by $\tilde{S}^2$.
 In these angular surfaces we now define two
new coordinates,
\begin{eqnarray}
 \theta = \int_{x}^{x_\mathrm{n}}{\cal{G}}^{-1/2}dx \:,
\hspace{1cm} \phi &=& z/\kappa \:,
                             \label{ang}
\end{eqnarray}
where $\phi$ ranges between $[0,2\pi]$ and $\kappa$ is an
arbitrary positive constant which will be discussed later. The
coordinate $\theta$ ranges between the north pole,
$\theta=\theta_\mathrm{n}=0$, and the south pole,
$\theta=\theta_\mathrm{s}$ (not necessarily at $\pi$). Rewritten
as a function of these new coordinates, the angular part of the
metric becomes $ds_{\tilde{S}^2}^2=d\theta^2+\kappa^2{\cal
G}\,d\phi^2$.  Note that when we set $A=0$ we have
$x_\mathrm{s}\leq x\leq x_\mathrm{n}$, with
$x_\mathrm{s}=-1$ and $x_\mathrm{n}=1$, $x=\cos \theta$, ${\cal
G}=1-x^2=\sin^2 \theta$, and $\kappa=1$. Therefore, in this case
the compact angular surface $\tilde{S}^2$ is a round $S^2$ sphere
which justifies the label given to the new angular coordinates.
When we set $A \neq 0$ the compact angular surface turns into a
deformed 2-sphere that we represent onwards by $\tilde{S}^2$.

Recently, the issue of the gravitational radiation emitted
by uniformly accelerated sources without horizons has been
analyzed by Bi\v c\'ak and Krtou\v s \cite{BicKrt} and the
radiative properties of accelerated black holes in dS background
have been analyzed by Krtou\v s and Podolsk\' y \cite{KrtPod}.

\subsection{\label{sec:Gen Nariai}The Nariai C-metric}
We will generate the Nariai C-metric from a
special extremal limit of the
dS C-metric. First we will describe this particular
near-extreme solution and then we will generate the Nariai
C-metric.

We are interested in a particular extreme dS C-metric, for which
the size of the acceleration horizon $y_A$ is equal to the size of
the outer charged horizon $y_+$. Let us label this degenerated
horizon by $\rho$, i.e.,  $y_A=y_+\equiv \rho$ with $\rho < y_-$.
 In this case, the function ${\cal F}(y)$ can be written as
\begin{eqnarray}
{\cal F}(y) =\frac{\rho^2-3\gamma}{\rho^4}
 (y-y_{\rm neg})(y-y_-)(y-\rho)^2\:,
 \label{Fextreme}
 \end{eqnarray}
where
\begin{eqnarray}
\gamma =\frac{\Lambda+3A^2}{3A^2}\:,
 \label{gamma}
 \end{eqnarray}
and the roots $\rho$, $y_{\rm neg}$ and $y_-$ are given by
\begin{eqnarray}
& & \rho =\frac{3m}{4q^2A}
 \left ( 1- \sqrt{1-\frac{8}{9}\frac{q^2}{m^2}} \:\right )
 \:,  \label{zerosy1} \\
& & y_{\rm neg} =\frac{\gamma \rho}{\rho^2-3\gamma}
 \left ( 1- \sqrt{\frac{\rho^2-2\gamma}{\gamma}} \:\right )
 \:, \label{zerosy2} \\
& & y_- =\frac{\gamma \rho}{\rho^2-3\gamma}
 \left ( 1+ \sqrt{\frac{\rho^2-2\gamma}{\gamma}} \:\right )
 \:.
 \label{zerosy3}
 \end{eqnarray}
The mass and the charge of the solution are written as functions
of $\rho$ as
\begin{eqnarray}
& &m =\frac{1}{A\rho}
 \left ( 1- \frac{2\gamma}{\rho^2} \right )
 \:, \nonumber \\
& & q^2 =\frac{1}{A^2\rho^2}
 \left ( 1- \frac{3\gamma}{\rho^2} \right )
 \:.
 \label{mq}
 \end{eqnarray}
The conditions $\rho < y_-$ and $q^2 \geq 0$ require that the
allowed range of $\rho$ is
\begin{eqnarray}
 \sqrt{3\gamma} \leq \rho<\sqrt{6\gamma} \:.
 \label{range-gamma}
\end{eqnarray}
The value of $y_-$ decreases monotonically with $\rho$ and we have
$\sqrt{6\gamma}<y_-<+\infty$. The mass and the charge of the
Nariai solution, denoted now as
$m_{\rm N}$ and $q_{\rm N}$ respectively,
are monotonically increasing functions
of $\rho$, and as we go from $\rho=\sqrt{3\gamma}$ into
$\rho=\sqrt{6\gamma}$ we have
\begin{eqnarray}
& &  \frac{1}{3}
  \frac{1}{\sqrt{\Lambda+3A^2}}\leq m_{\rm N}< \frac{\sqrt{2}}{3}
  \frac{1}{\sqrt{\Lambda+3A^2}}\:, \nonumber \\
& & 0 \leq q_{\rm N}< \frac{1}{2}
  \frac{1}{\sqrt{\Lambda+3A^2}}\:,
 \label{mq-cNariai}
 \end{eqnarray}
so the $A\neq 0$ extreme ($y_A=y_+$) solution has a lower maximum
mass and charge, and has a lower minimum mass than the
corresponding $A=0$ solution \cite{MelMosRom,MannRoss,BooMann},
and, for a fixed $\Lambda$, as the acceleration parameter $A$
grows these extreme values decrease monotonically. For a fixed
$\Lambda$ and for a fixed mass between $\sqrt{1/(9 \Lambda)}\leq
m<\sqrt{2/(9 \Lambda)}$, the allowed acceleration varies as
$\sqrt{1/(27m^2)-\Lambda/3} \leq A <\sqrt{2/(27m^2)-\Lambda/3}$.

We are now ready to obtain the Nariai C-metric.
 In order to generate it from the above
near-extreme dS C-metric we first set
\begin{eqnarray}
 y_A=\rho-\varepsilon, \:\:\:\:y_+=\rho+\varepsilon,
 \:\:\:\:\:\:{\rm with} \:\: \varepsilon<<1\:,
 \label{NariaiLimit}
\end{eqnarray}
in order that $\varepsilon$ measures the deviation from
degeneracy, and the limit $y_A\rightarrow y_+$ is obtained when
$\varepsilon \rightarrow 0$. Now, we introduce a new time
coordinate $\tau$ and a new radial coordinate $\chi$,
 \begin{eqnarray}
t= \frac{1}{\varepsilon {\cal K}}\,\tau \:, \:\:\:\:\:\:\:\:\:\:\:
y=\rho+\varepsilon \cos\chi \:,
 \label{NariaiCoord}
\end{eqnarray}
where
\begin{eqnarray}
{\cal K} = -\frac{\rho^2-3\gamma}{\rho^4} (\rho-y_{\rm
neg})(\rho-y_-) =\frac{2(\Lambda+3A^2)}{A^2\rho^2}-1\:, \nonumber
\\
& &
 \label{Kfactor}
\end{eqnarray}
and condition (\ref{range-gamma}) implies $0<{\cal K}\leq 1$ with
$q=0\Rightarrow {\cal K}=1$. In the limit $\varepsilon \rightarrow
0$, from (\ref{C-metric}) and (\ref{Fextreme}), we get the
gravitational field of the Nariai C-metric
\begin{eqnarray}
d s^2 &=& \frac{{\cal R}^2(x)}{{\cal K}} \left (-\sin^2\chi\,
d\tau^2 +d\chi^2\right ) \nonumber \\
& & + {\cal R}^2(x)\left
[{\cal G}^{-1}(x)dx^2+ {\cal G}(x)dz^2 \right ]
 \:,
 \label{Nariai-C-Metric}
\end{eqnarray}
where $\chi$ runs from $0$ to $\pi$ and
\begin{eqnarray}
& {\cal R}^2(x)=\left (Ax+\sqrt{\frac{2(\Lambda+3A^2)}{1+{\cal
K}}} \right )^{-2}\:,&
\nonumber \\
\!\!\!\!\!\!\!\!\!\!\!\!\!\!\!\!\!\!& \!\!\!\!\!\!\!\!\!
 {\cal G}(x) =1-x^2
 -\frac{2A}{3}\sqrt
{\frac{(1+{\cal K})(2-{\cal K})^2}{2(\Lambda+3A^2)}}\,x^3 -
\frac{A^2}{4}\frac{1-{\cal K}^2}{\Lambda+3A^2} \,x^4\,.&
 \nonumber \\
 & &
\label{Gfactor}
\end{eqnarray}
${\cal G}(x)$ has only two real roots, the south pole
$x_\mathrm{s}$ and the north pole $x_\mathrm{n}$. The angular
coordinate $x$ can range between these two poles, whose values are
calculated in Appendix \ref{sec:angular}. Under the coordinate
transformation (\ref{NariaiCoord}), the Maxwell field for the
magnetic case is still given by Eq. (\ref{F-mag}), while in the
electric case, (\ref{F-el-Lorentz}) becomes
 \begin{eqnarray}
 F=\frac{q}{{\cal K}}\,\sin \chi \,d\tau\wedge d\chi\:.
\label{F-el-Nariai}
\end{eqnarray}
 So, if we give the parameters $\Lambda$, $A$, and $q$ we can
construct the Nariai C-metric.
 The Nariai C-metric is
conformal to the topological product of two 2-dimensional
manifolds, $dS_2\times \tilde{S}^2$, with the conformal factor
${\cal R}^2(x)$ depending on the angular coordinate $x$, and
$\tilde{S}^2$ being a deformed 2-sphere.

 In order to obtain the $A=0$ limit, we first set
$\tilde{\rho}=A\rho$ [see Eq. (\ref{zerosy1})], a parameter that
has a finite and well-defined value when $A\rightarrow 0$. Then
when $A\rightarrow 0$ we have ${\cal K} \rightarrow {\cal
K}_0=2\Lambda/\tilde{\rho}^2-1$ and ${\cal R}^2(x) \rightarrow
{\cal R}_0^{\:2}=\tilde{\rho}^{-2}$, with ${\cal R}_0^{\:2}$ and
${\cal K}_0$ satisfying relations (\ref{relations}). This,
together with transformations (\ref{ang}), show that the Nariai
C-metric transforms into the Nariai solution (\ref{qNariai}) when
$A=0$.

 The limiting procedure that has been applied in this
subsection has generated a new exact solution that satisfies the
Maxwell-Einstein equations in a positive cosmological constant
background.

In order to give a physical interpretation to this extremal limit
of the dS C-metric, we first recall the physical interpretation of
the dS C-metric. This solution describes a pair of uniformly
accelerated black holes in a dS background, with the acceleration
being provided by the cosmological constant and by a strut between
the black holes that pushes them away or, alternatively, by a
string that connects and pulls the black holes in. The presence of
the strut or of the string is associated to the conical
singularities that exist in the C-metric (see, e. g.,
\cite{OscLem_dS-C}). Indeed, in general, if we draw a small circle
around the north or south pole, as the radius goes to zero, the
limit circunference/radius is not $2\pi$. There is a deficit angle
at the north pole given by (see \cite{OscLem_AdS-C,OscLem_dS-C}),
$\delta_\mathrm{{n}} =
 2\pi\left (1- \frac{\kappa}{2} |{\cal G}'(x_\mathrm{n})|\right )$
(where the prime means derivative with respect to $x$) and,
analogously, a similar conical singularity ($\delta_\mathrm{{s}}$)
is present at the south pole. The so far arbitrary parameter
$\kappa$ introduced in Eq. (\ref{ang}) plays its important role
here. Indeed, if we choose $\kappa^{-1}=\frac{1}{2}|{\cal
G}'(x_\mathrm{s})|$ we remove the conical singularity at the south
pole ($\delta_\mathrm{s}=0$). However, since we only have a single
constant $\kappa$ at our disposal and this has been fixed to avoid
the conical singularity at the south pole ($\delta_\mathrm{s}=0$),
we conclude that a conical singularity will be present at the
north pole with $\delta_\mathrm{n}<0$. This is associated to a
strut (since $\delta_\mathrm{n}<0$) that joins the two black holes
along their north poles and provides their acceleration. This
strut satisfies the relation $p=-\mu>0$, where $p$ and
$\mu=\delta_\mathrm{n}/(8\pi)$ are respectively its pressure and
its mass density (see \cite{OscLem_AdS-C}). There is another
alternative.  We can choose instead $\kappa^{-1}=\frac{1}{2}|{\cal
G}'(x_\mathrm{n})|$ and by doing so we avoid the deficit angle at
the north pole ($\delta_\mathrm{n}=0$), and leave a conical
singularity at the south pole ($\delta_\mathrm{s}>0$). This option
leads to the presence of a string (with $p=-\mu<0$) connecting the
black holes along their south poles that furnishes the
acceleration. Summarizing, when the conical singularity is at the
north pole, the pressure of the strut is positive, so it points
outwards and pushes the black holes apart, furnishing their
acceleration. When the conical singularity is at the south pole,
it is associated to a string between the two black holes with
negative pressure that pulls the black holes away from each other.

\begin{figure} [t]
\includegraphics*[height=2.7in]{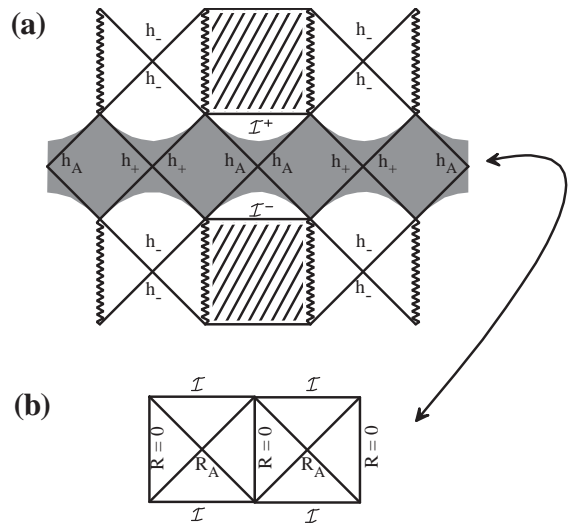}
\caption{\label{nariai-fig}
 (a) The whole figure is the Carter-Penrose
diagram of the dS C-metric.
 The shaded region represents the near-horizon area that, when the black hole
 horizon ($h_+$) approaches the acceleration horizon ($h_+$), gives the $dS_2$
 manifold of the Nariai C-metric solution. See discussion in the
 text.
 (b) Carter-Penrose diagram of the Nariai C-metric.
 }
\end{figure}

\begin{figure} [t]
\includegraphics*[height=2.4in]{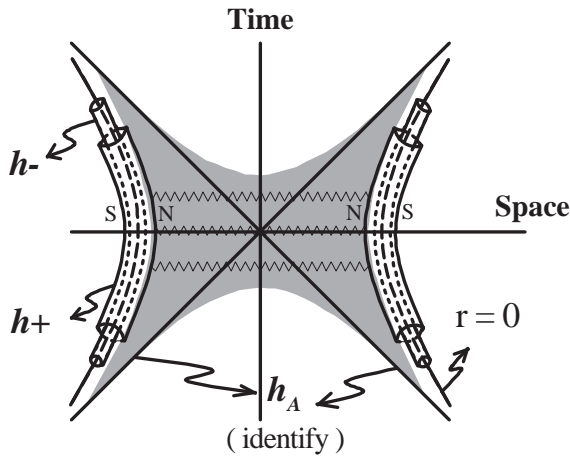}
\caption{\label{nariai2-fig}
 The whole figure represents schematically the two black holes of
 the dS C-metric approaching asymptotically the Rindler-like acceleration horizon.
 They are accelerated by a strut that joins them along their
 north poles.
 The shaded region represents the near-horizon area that, when the black hole
 horizon approaches the acceleration horizon, gives the Nariai C-metric
 solution (compare with Fig. \ref{nariai-fig}). The strut survives
 to the limiting process.
 }
\end{figure}

The causal structure of the dS C-metric has been analyzed in
detail in \cite{OscLem_dS-C}. The Carter-Penrose diagram of the
non-extreme charged dS C-metric is sketched in Fig.
\ref{nariai-fig}.(a) (whole figure) and has a structure that,
loosely speaking, can be divided into left, middle and right
regions. The middle region contains the null infinity, the past
infinity, ${\cal I^-}$, and the future infinity, ${\cal I^+}$, and
an accelerated Rindler-like horizon, $h_A$ (that coincides with
the cosmological horizon). The left and right regions both contain
a timelike curvature singularity (the zig-zag line), and an inner
($h_-$) and an outer ($h_+$) horizons associated to the charged
character of the solution. This diagram represents two
dS$-$Reissner-Nordstr\"{o}m black holes that approach
asymptotically the Rindler-like acceleration horizon (for a more
detailed discussion see \cite{OscLem_dS-C}). This is also
schematically represented in Fig. \ref{nariai2-fig} (whole
figure), where we explicitly show the strut that connects the two
black holes and provides their acceleration.

Now, as we have just seen, the Nariai C-metric can be
appropriately obtained from the vicinity of the black hole and
acceleration horizons in the limit in which the size of these two
horizons approach each other.  We now will see that the conical
singularity of the dS C-metric survives  the near-extremal
limiting procedure that generates the Nariai C-metric. Following
an elucidative illustration shown in \cite{MaldStrom} for the
Bertotti-Robinson solution (with $\Lambda<0$ and $A=0$), this
near-horizon region is sketched in Fig. \ref{nariai-fig}.(a) as a
shaded area, and from it we can identify some of the features of
the Nariai C-metric, e.g., the curvature singularity of the
original dS black hole is lost in the near-extremal limiting
procedure. But, more important, this shaded near-horizon region
also allows us to construct straightforwardly the Carter-Penrose
diagram of the Nariai C-metric drawn in Fig. \ref{nariai-fig}.(b).
The construction steps are as follows. First, as indicated by
(\ref{NariaiLimit}) and the second relation of
(\ref{NariaiCoord}), we let the cross lines that represent the
black hole horizon [$h_+$ in Fig. \ref{nariai-fig}.(a)] join
together with the cross lines that represent the acceleration
horizon [$h_A$ in Fig. \ref{nariai-fig}.(a)], and so on, i.e., we
do this joining ad infinitum with all the cross lines $h_+$ and
$h_A$. After this step all that is left from the original diagram
is a single cross line, i.e, all the spacetime that has originally
contained in the shaded area of Fig. \ref{nariai-fig}.(a) has
collapsed into two mutually perpendicular lines at $45^{\circ}$ at
$y=\rho$. Now, as indicated by the first relation of
(\ref{NariaiCoord}), when $\varepsilon \rightarrow 0$ the time
suffers an infinite blow up. To this blow up in the time
corresponds an infinite expansion in the Carter-Penrose diagram in
the vicinity of $y=\rho$. We then get again the shaded area of
Fig. \ref{nariai-fig}.(a), but now the cross lines of this shaded
area are all identified into a single horizon, and they no longer
have the original signature associated to $h_+$ and $h_A$ that
differentiated them. This is, in the shaded area of Fig.
\ref{nariai-fig}.(a) we must now erase the original labels $h_+$
and $h_A$. The Carter-Penrose diagram of the Nariai C-metric is
then given by Fig. \ref{nariai-fig}.(b), which is equivalent to
the diagram of the (1+1)-dimensional dS solution. Note that the
diagram of the $A=0$ dS$-$Reissner-Nordstr\"{o}m solution is
identical to the one of Fig. \ref{nariai-fig}.(a), as long as we
replace $h_A$ (acceleration horizon) by $h_c$ (cosmological
horizon). Applying the same construction process described just
above we find that the Carter-Penrose diagram of the Nariai
solution ($A=0$), described in subsection \ref{sec:Nariai}, is
also given by Fig. \ref{nariai-fig}.(b).

The Nariai near-horizon region is also sketched in Fig.
\ref{nariai2-fig} as a shaded area. This schematic figure is
clarifying in the sense that it indicates that the strut that
connects the two original black holes along their north pole
directions survives to the near-extremal limiting process and will
be present in the final result of the process, i.e., in the Nariai
C-metric. Indeed, note that the endpoints of the strut are at the
north pole of the event horizons of the two black holes and
crosses the acceleration horizon. As we saw just above, this
region suffers first a collapse followed by an infinite blow up
and during the process the strut is not lost. Thus the Nariai
C-metric (\ref{Nariai-C-Metric})-(\ref{Gfactor}) describes a
spacetime that is conformal to the product $dS_2\times
\tilde{S}^2$. To each point in the deformed 2-sphere corresponds a
$dS_2$ spacetime, except for one point which corresponds a $dS_2$
spacetime with an infinite straight strut. This strut has negative
mass density given by
\begin{equation}
 \mu =\frac{1}{4}{\biggl (}1- {\biggl |}\frac{{\cal G}'(x_\mathrm{{n}})}
 {{\cal G}'(x_\mathrm{{s}})}
 {\biggl |}{\biggr )}\:,
 \label{mass-density}
 \end{equation}
where ${\cal G}'(x)$ is the derivative of Eq. (\ref{Gfactor}), and
with a positive pressure $p=-\mu>0$. Alternatively, if we remove
the conical singularity at the north pole, the Nariai C-metric
describes a string with positive mass density
 $\mu =(1/4) \left ( 1- |{\cal G}'(x_\mathrm{{s}})/
 {\cal G}'(x_\mathrm{{n}})| \right )$
and negative pressure $p=-\mu<0$.

As we have said in subsection \ref{sec:Nariai}, the Nariai
solution ($A=0$) is unstable and, once created, it decays through
the quantum tunnelling process into a slightly non-extreme black
hole pair. We then expect that the Nariai C-metric is also
unstable and that it will decay into a slightly non-extreme pair
of black holes accelerated by a strut or by a string. The Nariai
C-metric instanton also plays an important role in the decay of
the dS space, since it can mediate the Schwinger-like quantum
process of pair creation of black holes in a dS background
\cite{OscLem-PCdS}. Indeed, as we said in subsection
\ref{sec:Nariai}, the Nariai instanton ($A=0$) has been used
\cite{MelMosRom,MannRoss,BooMann,VolkovWipf} to study the pair
creation of dS black holes materialized and accelerated by the
cosmological background field. Moreover, the Euclidean ``Nariai"
flat C-metric \cite{flatPC} and Ernst solution \cite{ErnstPC}
(also discussed in this paper in section \ref{sec:dS C-metric})
have been used to analyze the process of pair production of
$\Lambda=0$ black holes, accelerated by a string or by an
electromagnetic external field, respectively. Therefore, it is
natural to expect that the Euclidean Nariai limit of the dS
C-metric mediates the process of pair creation of black holes in a
cosmological background, that are then accelerated by a string, in
addition to the cosmological background field. The picture would
be that of the nucleation, in a dS background, of a Nariai C
universe, whose string then breaks down and a pair of dS black
holes is created at the endpoints of the string. This expectation
is confirmed in \cite{OscLem-PCdS}.

\subsection{\label{sec:Gen BR dS}The Bertotti-Robinson dS C-metric}

Now, we are interested in another particular extreme dS C-metric
(usually called cold solution when $A=0$
\cite{MelMosRom,MannRoss,BooMann}) for which the size of the outer
charged black hole horizon $y_+$ is equal to the size of the inner
charged horizon $y_-$. Let us label this degenerated horizon by
$\rho$, such that, $y_+=y_-\equiv \rho$ and $\rho > y_A$. This
solution requires the presence of the electromagnetic charge. In
this case, the function ${\cal F}(y)$ can be written as
\begin{eqnarray}
{\cal F}(y) =\frac{\rho^2-3\gamma}{\rho^4}
 (y-y_{\rm neg})(y-y_A)(y-\rho)^2\:,
 \label{Fextreme-br}
 \end{eqnarray}
with $\gamma$ given by Eq. (\ref{gamma}), the roots $\rho$ and
$y_{\rm neg}$ are defined by Eqs. (\ref{zerosy1}) and
(\ref{zerosy2}), respectively, and $y_A$ is given by
\begin{eqnarray}
 y_A =\frac{\gamma \rho}{\rho^2-3\gamma}
 \left ( 1+ \sqrt{\frac{\rho^2-2\gamma}{\gamma}} \:\right )
 \:.
 \label{zerosy3-br}
 \end{eqnarray}
Eq. (\ref{mq}) defines the mass and the charge of the solution as
a function of $\rho$, and, for a fixed $A$ and $\Lambda$, the
ratio $q/m$ is higher than $1$. The conditions $\rho> y_A$ and
$q^2 > 0$ require that the allowed range of $\rho$ is
\begin{eqnarray}
\rho>\sqrt{6\gamma} \:.
 \label{range-gamma-br}
\end{eqnarray}
The value of $y_A$ decreases monotonically with $\rho$ and we have
$\sqrt{\gamma}<y_A<\sqrt{6\gamma}$. Contrary to the Nariai case,
the mass and the charge of the Bertotti-Robinson solution,
$m_{\rm BR}$ and $q_{\rm BR}$, respectively,
are monotonically
decreasing functions of $\rho$, and as we come from $\rho=+\infty$
into $\rho=\sqrt{6\gamma}$ we have
\begin{eqnarray}
& &  0< m_{\rm BR}< \frac{\sqrt{2}}{3}
  \frac{1}{\sqrt{\Lambda+3A^2}}\:, \nonumber \\
& & 0< q_{\rm BR}< \frac{1}{2}
  \frac{1}{\sqrt{\Lambda+3A^2}}\:,
 \label{mq-cold}
 \end{eqnarray}
so the $A\neq 0$ extreme ($y_+=y_-$) solution has a lower maximum
mass and charge than the corresponding $A=0$ solution
\cite{MelMosRom,MannRoss,BooMann} and, for a fixed $\Lambda$, as
the acceleration parameter $A$ grows this maximum value decreases
monotonically. For a fixed $\Lambda$ and for a fixed mass below
$\sqrt{2/(9 \Lambda)}$, the maximum value of the acceleration is
$\sqrt{2/(27m^2)-\Lambda/3}$.

We are now ready to generate the Bertotti-Robinson dS C-metric
from the above cold dS C-metric. We first set
\begin{eqnarray}
 y_+=\rho-\varepsilon, \:\:\:\:y_-=\rho+\varepsilon,
 \:\:\:\:\:\:{\rm with} \:\: \varepsilon<<1\:,
 \label{NariaiLimit-br}
\end{eqnarray}
in order that $\varepsilon$ measures the deviation from
degeneracy, and the limit $y_+\rightarrow y_-$ is obtained when
$\varepsilon \rightarrow 0$. Now, we introduce a new time
coordinate $\tau$ and a new radial coordinate $\chi$,
 \begin{eqnarray}
t= \frac{1}{\varepsilon {\cal K}}\,\tau \:, \:\:\:\:\:\:\:\:\:\:\:
y=\rho+\varepsilon \cosh\chi \:,
 \label{NariaiCoord-br}
\end{eqnarray}
where
\begin{eqnarray}
{\cal K} =\frac{\rho^2-3\gamma}{\rho^4}
 (\rho-y_{\rm neg})(\rho-y_A)
= 1-\frac{2(\Lambda+3A^2)}{A^2\rho^2}\:, \nonumber \\
& &
 \label{Kfactor-br}
\end{eqnarray}
and condition (\ref{range-gamma-br}) implies $0<{\cal K}<1$. In
the limit $\varepsilon \rightarrow 0$, from (\ref{C-metric}) and
(\ref{Fextreme-br}), the metric becomes
\begin{eqnarray}
d s^2 &=& \frac{{\cal R}^2(x)}{{\cal K}}
\left (-\sinh^2\chi\, d\tau^2 +d\chi^2\right ) \nonumber \\
&+&
 {\cal R}^2(x)\left [{\cal G}^{-1}(x)dx^2+ {\cal G}(x)dz^2 \right ]
 \:,
 \label{br-ds-C-Metric}
\end{eqnarray}
where $\chi$ is unbounded and
\begin{eqnarray}
 &{\cal R}^2(x)=
\left(Ax+\sqrt{\frac{2(\Lambda+3A^2)}{1-{\cal K}}}\right )^{-2}
\:, & \nonumber \\
 & {\cal G}(x) = 1-x^2
 -\frac{2A}{3}
\sqrt{\frac{(1-{\cal K})(2+{\cal K})^2}{2(\Lambda+3A^2)}}\,x^3
  -\frac{A^2}{4}\frac{1-{\cal K}^2}{\Lambda+3A^2} \,x^4\:.& \nonumber \\
& &
 \label{Gfactor-br}
\end{eqnarray}
${\cal G}(x)$ has only two real roots, the south pole
$x_\mathrm{s}$ and the north pole $x_\mathrm{n}$. The angular
coordinate $x$ can range between these two poles whose value is
calculated in appendix \ref{sec:angular}. Under the coordinate
transformation (\ref{NariaiCoord-br}), the Maxwell field for the
magnetic case is still given by Eq. (\ref{F-mag}), while in the
electric case, (\ref{F-el-Lorentz}) becomes
 \begin{eqnarray}
 F=-\frac{q}{{\cal K}}\,\sinh \chi \,d\tau\wedge d\chi\:.
 \label{F-el-br}
\end{eqnarray}
So, if we give the parameters $\Lambda$, $A$, and $q$ we can
construct the Bertotti-Robinson dS C-metric.  This solution is
conformal to the topological product of two 2-dimensional
manifolds, $AdS_2\times \tilde{S}^2$, with the conformal factor
${\cal R}^2(x)$ depending on the angular coordinate $x$.

In order to obtain the $A=0$ limit, we first set
$\tilde{\rho}=A\rho$ [see Eq. (\ref{zerosy1})], a parameter that
has a finite and well-defined value when $A\rightarrow 0$. Then
when $A\rightarrow 0$ we have ${\cal K} \rightarrow {\cal
K}_0=1-2\Lambda/\tilde{\rho}^2$ and ${\cal R}^2(x) \rightarrow
{\cal R}_0^{\:2}=\tilde{\rho}^{-2}$. This, together with
transformations (\ref{ang}), show that the Bertotti-Robinson dS
C-metric transforms into the dS counterpart of the
Bertotti-Robinson solution discussed in subsection \ref{sec:BR},
when $A=0$. The limiting procedure that has been applied in this
subsection has generated a new exact solution that satisfies the
Maxwell-Einstein equations in a positive cosmological constant
background.

\begin{figure} [t]
\includegraphics*[height=2.0in]{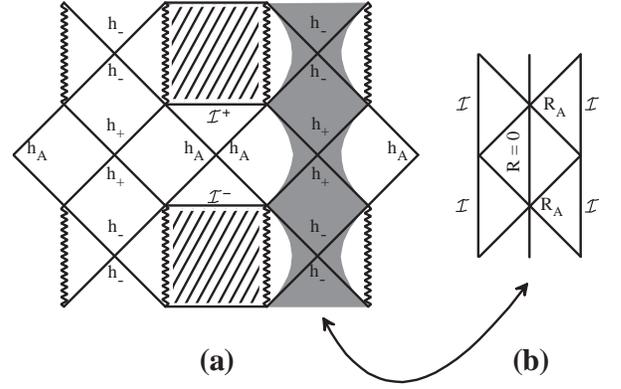}
\caption{\label{br-fig}
 (a) The whole figure is the Carter-Penrose diagram of the dS C-metric.
 The shaded region represents the near-horizon area that, when the
inner black hole
 horizon ($h_-$) approaches the outer black hole horizon ($h_+$),
gives the $AdS_2$
 manifold of the Bertotti-Robinson C-metric. See discussion in the
 text.
  (b) Carter-Penrose diagram of the Bertotti-Robinson C-metric.
 }
\end{figure}

We have just seen that the Bertotti-Robinson dS C-metric can be
appropriately obtained from the vicinity of the inner and outer
black hole horizons in the limit in which the size of these two
horizons approach each other. This near-horizon region is sketched
in Fig. \ref{br-fig}.(a) as a shaded area, and from it we can
construct straightforwardly the Carter-Penrose diagram of the
Bertotti-Robinson dS C-metric drawn in Fig. \ref{br-fig}.(b). The
construction steps are as follows. First, as indicated by
(\ref{NariaiLimit-br}) and the second relation of
(\ref{NariaiCoord-br}), we let the cross lines that represent the
black hole Cauchy  horizon [$h_-$ in Fig. \ref{br-fig}.(a)] join
together with the cross lines that represent the  black hole event
horizon [$h_+$ in Fig. \ref{br-fig}.(a)], and so on, i.e., we do
this junction ad infinitum with all the cross lines $h_-$ and
$h_+$. After this step all that is left from the original diagram
is a single cross line, i.e, all the spacetime that has originally
contained in the shaded area of Fig. \ref{br-fig}.(a) has
collapsed into two mutually perpendicular lines at $45^{\circ}$ at
$y=\rho$. Now, as indicated by the first relation of
(\ref{NariaiCoord-br}), when $\varepsilon \rightarrow 0$ the time
suffers an infinite blow up. To this blow up in the time
corresponds an infinite expansion in the Carter-Penrose diagram in
the vicinity of $y=\rho$ that generates an $AdS_2$ region. We then
get again the shaded area of Fig. \ref{br-fig}.(a), but now the
cross lines of this shaded area are all identified into a single
line, and they no longer have the original lables associated to
$h_-$ and $h_+$ that differentiated them. This is, in the shaded
area of Fig. \ref{br-fig}.(a) we must now erase the original
labels $h_-$ and $h_+$. The Carter-Penrose diagram of the
Bertotti-Robinson dS C-metric is then given by Fig.
\ref{br-fig}.(b), which is equivalent to the diagram of the
(1+1)-dimensional AdS solution. Note that the diagram of the $A=0$
Reissner-Nordstr\"{o}m$-$dS solution is identical to the one of
Fig. \ref{br-fig}.(a). Therefore, applying the same construction
process described just above we find that the Carter-Penrose
diagram of the Bertotti-Robinson dS solution ($A=0$) is similar to
the one of Fig. \ref{br-fig}.(b). The diagram of the
Bertotti-Robinson solution with $\Lambda=0$ described by
(\ref{br}) is also given by  Fig. \ref{br-fig}.(b).

\begin{figure} [t]
\includegraphics*[height=2.0in]{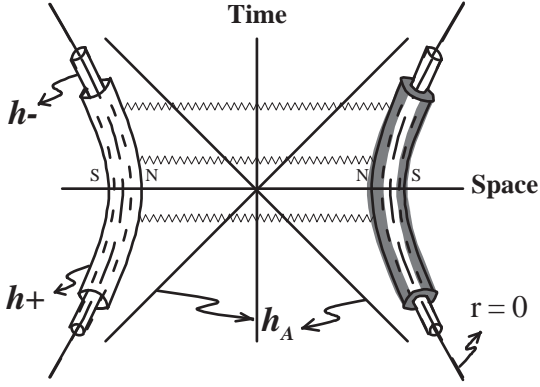}
\caption{\label{br2-fig}
 The whole figure represents schematically the two accelerated black holes of
 the dS C-metric.
 The shaded region represents the near-horizon area that, when the
inner black hole
 horizon approaches the outer black hole horizon, gives
 the Bertotti-Robinson C-metric (compare with Fig. \ref{br-fig}).
The strut does not
 survive to the limiting process.
 }
\end{figure}

The Bertotti-Robinson near-horizon region is also sketched in Fig.
\ref{br2-fig} as a shaded area. This schematic figure is
clarifying in the sense that it indicates that the strut that
connects the two original black holes along their north pole
directions does not survive to the near-extremal limiting process
and will not be present in the final result of the process, i.e.,
in the Bertotti-Robinson dS C-metric. However, a reminiscence of
this strut remains in the final solution. Indeed, the angular
factor of the Bertotti-Robinson dS C-metric [which, remind,
describes a deformed 2-sphere $\tilde{S}^2$ with a fixed size
${\cal R}^2(x)$ given by  (\ref{Gfactor-br})] has a conical
singularity at least at one of its poles. We can choose the
parameter $\kappa$, introduced in (\ref{ang}), in order to have a
conical singularity only at the north pole
($\delta_\mathrm{s}=0$), or only at the south pole
($\delta_\mathrm{n}=0$), however we cannot eliminate both. When
the parameter $A$ is set to zero, the conical singularities
disappear, the angular factor describes a round 2-sphere with
fixed radius, and the Bertotti-Robinson dS C-metric reduces into
the Bertotti-Robinson dS solution with topology $AdS_2 \times
S^2$.

\subsection{\label{sec:Gen Nariai-BR dS}The Nariai Bertotti-Robinson
dS C-metric}
\begin{figure} [t]
\includegraphics*[height=2.1in]{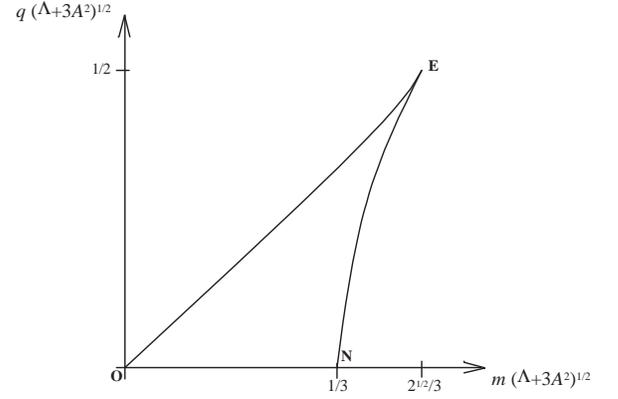}
\caption{\label{nbr-mq-fig}
 Relation $m\sqrt{\Lambda+3A^2}\times
q\sqrt{\Lambda+3A^2}$, for a fixed value of $A$ and $\Lambda$, for
the extremal limits of the dS C-metric. $NE$ represents the Nariai
C-metric (with point  $N$ representing the neutral case), $OE$
represents the Bertotti-Robinson dS C-metric, and point $E$
represents the Nariai Bertotti-Robinson dS C-metric.
 }
\end{figure}

As the previous sections and the label suggest, the Nariai
Bertotti-Robinson dS C-metric will be generated  from the extremal
limit of a very particular dS C-metric (usually called ultracold
solution  when $A=0$ \cite{MelMosRom,MannRoss,BooMann}) for which
the size of the three horizons ($y_A$, $y_+$ and $y_-$) are equal,
and let us label this degenerated horizon by $\rho$:
$y_A=y_+=y_-\equiv \rho$.
 In this case, the function ${\cal F}(y)$ is given by
Eq. (\ref{Fextreme}) with $y_-=\rho$, and $\gamma$ defined in Eq.
(\ref{gamma}). The negative root $y_{\rm neg}$ is given by Eq.
(\ref{zerosy2}) and
\begin{eqnarray}
\rho=\sqrt{6\gamma} \:.
 \label{range-gamma-N-br}
\end{eqnarray}
The mass and the charge
of the Nariai Bertotti-Robinson dS C-metric solution ,
$m_{\rm NBR}$ and $q_{\rm NBR}$, respectively,
are given by
\begin{eqnarray}
& &m_{\rm NBR} =\frac{\sqrt{2}}{3}
 \sqrt{\frac{1}{\Lambda+3A^2}}
 \:, \nonumber \\
& & q_{\rm NBR} =\frac{1}{2}\sqrt{\frac{1}{\Lambda+3A^2}}
 \:,
 \label{mq-N-br}
\end{eqnarray}
and these values are the maximum values of the mass and charge of
both the Nariai C and Bertotti-Robinson C solutions, Eqs.
(\ref{mq-cNariai}) and (\ref{mq-cold}), respectively. To clarify
the nature of these solutions, a diagram
$m\sqrt{\Lambda+3A^2}\times q\sqrt{\Lambda+3A^2}$ is plotted in
Fig. \ref{nbr-mq-fig}. For a fixed value of $A$ and $\Lambda$, the
allowed range of the mass and charge of the Nariai C-metric, of
the Bertotti-Robinson dS C-metric, and of the Nariai
Bertotti-Robinson dS C-metric is shown.

We are now ready to generate the Nariai Bertotti-Robinson dS
C-metric from the above ultracold dS C-metric. We first set
\begin{eqnarray}
 \rho=\sqrt{6\gamma}-\varepsilon, \:\:\:\:y_-=\sqrt{6\gamma}+\varepsilon \:,
 \:\:\:\:\:\:{\rm with} \:\: \varepsilon<<1\:.
 \label{NariaiLimit-N-br}
\end{eqnarray}
Now, we introduce a new time coordinate $\tau$ and a new radial
coordinate $\chi$,
 \begin{eqnarray}
t= \frac{1}{2 \varepsilon^2 {\cal K}}\,\tau \:,
\:\:\:\:\:\:\:\:\:\:\: y=\sqrt{6\gamma}+
 \varepsilon \cosh(\sqrt{2\varepsilon{\cal K}}\:\chi) \:,
 \label{NariaiCoord-N-br}
\end{eqnarray}
where
\begin{eqnarray}
{\cal K} &=& \frac{\rho^2-3\gamma}{\rho^4}
 (\rho-y_{\rm neg}) = \frac{1}{3}
 \sqrt{\frac{2A^2}{\Lambda+3A^2}}\:.
 \label{Kfactor-N-br}
\end{eqnarray}
 In the limit $\varepsilon \rightarrow
0$ the metric (\ref{C-metric}) becomes
\begin{eqnarray}
d s^2 \!\!&=&\!\! {\cal R}^2(x) \left [-\chi^2\, d\tau^2 +d\chi^2
 +{\cal G}^{-1}(x)dx^2+ {\cal G}(x)dz^2 \right ],
 \nonumber \\
 \label{Nariai-C-Metric-N-br}
\end{eqnarray}
with
\begin{eqnarray}
& {\cal R}^2(x) = \left (Ax+\sqrt{2(\Lambda+3A^2)}\right )^{-2}\:, &
\nonumber \\
 &  {\cal G}(x) = 1-x^2
 -\frac{2A}{3}\sqrt{\frac{2}{\Lambda+3A^2}}\,x^3
 - \frac{A^2}{4(\Lambda+3A^2)} \,x^4\:.&
 \label{Gfactor-N-br}
\end{eqnarray}
${\cal G}(x)$ has only two real roots, the south pole
$x_\mathrm{s}$ and the north pole $x_\mathrm{n}$. The angular
coordinate $x$ can range between these two poles whose value is
calculated in appendix \ref{sec:angular}. Notice that the
spacetime factor $-\chi^2\, d\tau^2 +d\chi^2$ is just
${\mathbb{M}}^{1,1}$ in Rindler coordinates. Therefore, under the
usual coordinate transformation
$\chi=\sqrt{\bar{x}^2-\bar{t}^{\,2}}$ and $\tau={\rm
arctanh(\bar{t}/\bar{x})}$, this factor transforms into
$-d\bar{t}^{\,2} +d\bar{x}^2$.
 Under the coordinate transformation (\ref{NariaiCoord-N-br}), the
Maxwell field for the magnetic case is still given by Eq.
(\ref{F-mag}), while in the electric case, (\ref{F-el-Lorentz})
becomes
 \begin{eqnarray}
 F=-q\,\chi \,d\tau\wedge d\chi\:.
 \label{F-el-N-br}
\end{eqnarray}
The Nariai Bertotti-Robinson dS C-metric is conformal to the
topological product of two 2-dimensional manifolds,
${\mathbb{M}}^{1,1}\times \tilde{S}^2$, with the conformal factor
${\cal R}^2(x)$ depending on the angular coordinate $x$.

In the $A=0$ limit, ${\cal R}^2(x) \rightarrow (2\Lambda)^{-1}$,
and one obtains the Nariai Bertotti-Robinson solution
$ds^2=(2\Lambda)^{-1}(-d\bar{t}^{\,2} +d\bar{x}^2+ d\theta^2+
\sin^2\theta\,d\phi^2)$, which has the topology
${\mathbb{M}}^{1,1}\times S^2$ \cite{MelMosRom,MannRoss,BooMann}).

\begin{figure} [t]
\includegraphics*[height=1.7in]{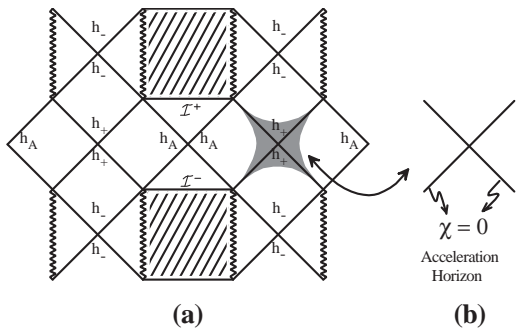}
\caption{\label{n-br-fig}
 (a) The whole figure is the Carter-Penrose diagram of the dS C-metric.
 The shaded region represents the near-horizon area that, when the
 inner black hole horizon ($h_-$) and the acceleration horizon ($h_A$)
 approach the outer black hole horizon ($h_+$), gives the Rindler manifold of
 the Nariai Bertotti-Robinson C-metric. This shaded region is the
 intersection of the shaded areas of Figs. \ref{nariai-fig}.(a) and
 \ref{br-fig}.(a). See discussion in the text.
 (b) Kruskal diagram of the Nariai Bertotti-Robinson dS C-metric.
 }
\end{figure}

We have just seen that the Nariai Bertotti-Robinson dS C-metric
can be appropriately obtained from the vicinity of the accelerated
and black hole horizons in the limit in which the size of these
three horizons approach each other. This near-horizon region is
sketched in Fig. \ref{n-br-fig}.(a) as a shaded area, and can be
viewed as the intersection of the shaded areas of Figs.
\ref{nariai-fig}.(a) and \ref{br-fig}.(a). From it we can
construct straightforwardly, following the construction steps
already sketched in subsections \ref{sec:Gen Nariai} and
\ref{sec:Gen BR dS}, the Kruskal diagram of the Nariai
Bertotti-Robinson dS C-metric drawn in Fig. \ref{n-br-fig}.(b).
This diagram is equivalent to the Kruskal diagram of the Rindler
solution. The strut that connects the two original black holes
along their north pole directions  survives to the near-extremal
limiting process. Thus the Nariai Bertotti-Robinson dS C-metric
describes a spacetime that is conformal to the product
${\mathbb{M}}^{1,1} \times \tilde{S}^2$. To each point in the
deformed 2-sphere corresponds a ${\mathbb{M}}^{1,1}$ spacetime,
except for one point which corresponds a ${\mathbb{M}}^{1,1}$
spacetime with an infinite straight strut or string, with a mass
density and pressure satisfying $p=-\mu$. In an analogous way to
the one that occurs with the Nariai C universe (see section
\ref{sec:Gen Nariai L=0}), we expect that the Nariai
Bertotti-Robinson dS C universe is unstable and, once created, it
decays through the quantum tunnelling process into a slightly
non-extreme black hole pair. The picture would be that of the
nucleation, in a dS background, of a Nariai Bertotti-Robinson dS C
universe, whose string then breaks down and a pair of dS black
holes is created at the endpoints of the string. This expectation
is confirmed in \cite{OscLem-PCdS}. When the parameter $A$ is set
to zero, the conical singularities disappear (and so the strut or
string are no longer present) and the angular factor describes a
round 2-sphere with fixed radius, and the topology of the Nariai
Bertotti-Robinson dS C-metric reduces into
${\mathbb{M}}^{1,1}\times S^2$.

\section{\label{sec:ExtLim L=0}Extremal limits of the flat C-metric
and of the Ernst solution}
The Euclidean version of the ``Nariai"  flat C-metric that will be
discussed in subsection \ref{sec:Gen Nariai L=0} has already been
used previously \cite{flatPC,ErnstPC,DGKT} in the study of the
quantum process of pair creation of black holes. However, as far
as we know, the Bertotti-Robinson  flat C-metric (discussed in
subsection \ref{sec:Gen BR L=0}) and the Nariai Bertotti-Robinson
flat C-metric (discussed in subsection \ref{sec:Gen Nariai-BR
L=0}) have not been written explicitly.

\subsection{\label{sec:L=0 C-metric}The  flat C-metric and the Ernst solution}

\subsubsection{\label{sec:sub L=0 C-metric}The  flat C-metric}
\begin{figure} [t]
\includegraphics*[height=1.6in]{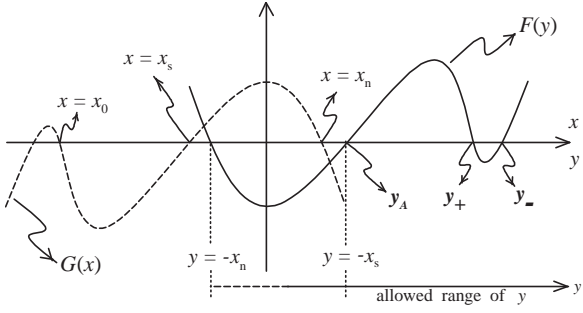}
\caption{\label{g3-flat} Shape of ${\cal G}(x)$ and ${\cal F}(y)$
for a general non-extremal charged massive flat C-metric studied
in section \ref{sec:ExtLim L=0}. The allowed range of $x$ is
between $x_\mathrm{s}$ and $x_\mathrm{n}$ where ${\cal G}(x)$ is
positive and compact. The permitted range of $y$ is $-x\leq y <
+\infty$. The presence of an accelerated horizon is indicated by
$y_A$ and the inner and outer charged horizons by $y_-$ and $y_+$.
In the ``Nariai" case, $y_+$ and $y_A$ coincide (see section
\ref{sec:Gen Nariai L=0}), in the Bertotti-Robinson case, $y_-$
and $y_+$ coincide (see section \ref{sec:Gen BR L=0}), and in the
Nariai Bertotti-Robinson case, $y_-$, $y_+$ and $y_A$ coincide
(see section \ref{sec:Gen Nariai-BR L=0}). Note that when $y_+$
and $y_A$ coincide, the same occurs with $x_\mathrm{s}$ and
$x_0$.}
\end{figure}

The gravitational field and the electromagnetic field of the
massive charged flat C-metric is given by Eqs.
(\ref{C-metric})-(\ref{F-el-Lorentz})  (see \cite{KW}), as long as
we set $\Lambda=0$ in Eq. (\ref{FG}), and therefore we have now
${\cal F}(y)=-{\cal G}(-y)$ [see Fig. \ref{g3-flat} for the
general shape of ${\cal F}(y)$ and ${\cal G}(x)$]. The discussion
of subsection \ref{sec:dS C-metric} applies then directly also to
the $\Lambda=0$ case. In particular, in general the flat C-metric
has a conical singularity at one of its poles that is interpreted
\cite{KW} as a strut joining the north poles of the two black
holes that accelerates them apart, or alternatively, as two
strings from infinity into the south pole of each one of the black
holes that that pushes them into infinity. For the actual issue of
the gravitational radiative properties of accelerated black holes
described by the flat C-metric see Bi\v c\'ak, and Pravda and
Pravdova \cite{BPP}.

\subsubsection{\label{sec:sub Ernst}The  Ernst solution}
Ernst \cite{Ernst} has employed a Harrison-type transformation to
the charged flat C-metric in order to append a suitably chosen
external electromagnetic field. With this procedure the Ernst
solution is free of conical singularities at both poles and the
acceleration that drives away the two oppositely charged
Reissner-Nordstr\"{o}m black holes is provided by the external
electromagnetic field.  The gravitational field of the Ernst
solution is \cite{Ernst}
\begin{eqnarray}
 d s^2 = \frac{\Sigma^2 ( -{\cal F}dt^2+
 {\cal F}^{-1}dy^2+{\cal G}^{-1}dx^2
 +  \Sigma^{-4} {\cal G}\,dz^2 )} {[A(x+y)]^2} \:, \nonumber\\
 \label{Ernst}
 \end{eqnarray}
 where ${\cal F}(y)$ and ${\cal G}(x)$ are given by (\ref{FG}) with
$\Lambda=0$, and
 \begin{eqnarray}
 \Sigma(x,y)={\biggl (}1+\frac{1}{2}\, q \, {\cal E}_0\, x{\biggr )}^2
   + \frac{{\cal E}_0^2 \, {\cal G}(x)}{4A^2 \,(x+y)^2}\:.
 \label{factor}
 \end{eqnarray}
In the electric solution one has $q\equiv e$ and ${\cal E}_0\equiv
E_0$, i.e., $q$ and ${\cal E}_0$  are  respectively the electric
charge and the external electric field. In the magnetic solution
one has $q\equiv g$ and ${\cal E}_0\equiv B_0$, i.e., $q$ and
${\cal E}_0$ are respectively the magnetic charge and the external
magnetic field. The electromagnetic potential of the magnetic
Ernst solution is \cite{Ernst}
\begin{eqnarray}
 A_{z}=-\frac{2}{\Sigma \, B_0 }{\biggl (}1
+\frac{1}{2}\, g \,B_0\, x{\biggr )}\:.
 \label{pot-Ernst-B}
\end{eqnarray}
while for the electric Ernst solution it is given by \cite{Brown}
\begin{eqnarray}
A_t &=& qy -\frac{E_0}{2A^2} \frac{{\cal F}(y)}{(x+y)^2}
 \left ( 1+e \, E_0 \, x-\frac{1}{2} e \, E_0 \, y \right )
 \nonumber \\
 &-&\frac{E_0}{2A^2}(1+r_- A y)(1+r_+ A y) (1-e\,E_0\,y/2)
\:,  \nonumber \\
& &
\label{pot-Ernst-E}
\end{eqnarray}
with $r_+ r_-=e^2$ and $r_+ +r_-=2m$. This exact solution
describes two oppositely charged Reissner-Nordstr\"{o}m black
holes accelerating away from each other in a magnetic Melvin or in
an electric Melvin-like background, respectively.

Technically, the Harrison-type transformation employed to generate
the Ernst solution introduces, in addition to the parameter $\kappa$,
a new parameter, the external field
${\cal E}_0$ that when
appropriately chosen allow us to eliminate the conical
singularities at both poles. Indeed, applying a procedure
analogous to the one employed in section \ref{sec:dS C-metric},
but now focused in spacetime (\ref{Ernst}), one introduces the new
angular coordinates
\begin{eqnarray}
 \tilde{\theta} = \int_{x_\mathrm{n}}^{x}\tilde{\cal{G}}^{-1/2}dx \:,
                        \:\:\:\:\:\: \:\:\:\:\:\:
\tilde{\phi} = z/\tilde{\kappa} \:,
                             \label{ang-Ernst}
\end{eqnarray}
where $\tilde{\cal{G}}(x,y)=\Sigma^{-2}(x,y){\cal{G}}(x)$. As
before, the conical singularity at the south pole is avoided by
choosing
\begin{equation}
\tilde{\kappa}^{-1}=\frac{1}{2}\left | \tilde{\cal
G}'(x_\mathrm{s}) \right | \,,
  \label{k-s-Ernst}
 \end{equation}
while the conical singularity at the north pole can now be also
eliminated by choosing the value of the external field
 ${\cal E}_0$ to satisfy
\begin{equation}
 \left | \tilde{\cal G}'(x_\mathrm{n}) \right | =
  \left | \tilde{\cal G}'(x_\mathrm{s}) \right | \:.
  \label{k=Ernst}
\end{equation}
An interesting support to the physical interpretation given to the
Ernst solution is the fact that in the particle limit, i.e.,  for
small values of $mA$, the condition (\ref{k=Ernst}) implies the
classical Newton's law \cite{Ernst}
\begin{equation}
q \, {\cal E}_0 \approx m \, A\:. \label{Lorentz}
\end{equation}
So, in this regime, the acceleration is indeed provided by the
Lorentz force. For a more detailed discussion on the properties of
the C-metric see \cite{DGKT,ErnstPC}.
Note that in a cosmological constant
background  the Harrison transformation does not work, since it
does not leave invariant the
cosmological term in the action. Thus Ernst's trick cannot be used.

\subsection{\label{sec:Gen Nariai L=0}The ``Nariai"  flat C-metric}
The Nariai solution \cite{Nariai} is originally a solution in the
$\Lambda>0$ background which can be obtained from a near-extremal
limit of the dS black hole when the outer black horizon approaches
the cosmological horizon. Therefore, it would seem not appropriate
to use the name Nariai to label a $\Lambda=0$ solution. However,
in the flat C-metric the acceleration horizon plays the role of
the cosmological horizon. Moreover, the limit $A=0$ of the
solution discussed in this subsection is equal to the limit
$\Lambda=0$ of the Nariai solution (see Appendix \ref{sec:lim L=0
Nariai}). In this context, we find appropriate to label this
solution by ``Nariai" in between commas, in order to maintain the
nomenclature of the paper.

This solution can be obtained from a near-extremal limit of the
flat C-metric when the outer black horizon approaches the
acceleration horizon. This is the way it has been first
constructed in \cite{DGKT,ErnstPC}. However, given the Nariai
C-metric ($\Lambda>0$) generated in subsection \ref{sec:Gen
Nariai}, we can construct the ``Nariai"  flat C-metric
($\Lambda=0$) by taking  directly the $\Lambda=0$ limit of Eqs.
(\ref{Nariai-C-Metric})-(\ref{Gfactor}). The gravitational field
of the ``Nariai"  flat C-metric is then given by Eq.
(\ref{Nariai-C-Metric}) with
\begin{eqnarray}
& {\cal R}^2(x) = A^{-2}\left (x+\sqrt{\frac{6}{1+{\cal K}}}\right
)^{-2}\:,&
\nonumber \\
 & {\cal G}(x) = 1-x^2
 -\sqrt{\frac{2(1+{\cal K})(2-{\cal K})^2}{27}}\,x^3
- \frac{1-{\cal K}^2}{12} \,x^4\:, &
 \label{Gfactor L=0}
\end{eqnarray}
where $0<{\cal K}\leq 1$ and ${\cal K}=1$ when $q=0$. ${\cal
G}(x)$ has only two real roots, the south pole $x_\mathrm{s}$ and
the north pole $x_\mathrm{n}$. The angular coordinate $x$ can
range between these two poles whose value is calculated in
Appendix \ref{sec:angular}.

There is a great difference between the Nariai C solution of the
$\Lambda>0$ case and the ``Nariai'' C solution of the $\Lambda=0$
case. In subsection \ref{sec:Gen Nariai} we saw that the Nariai
C-metric ($\Lambda>0$) has a conical singularity at one of the
poles of its deformed 2-sphere. This feature is no longer present
in the ``Nariai"  flat C-metric, i.e., it is free of conical
singularities. Indeed, in the flat C-metric we have ${\cal
F}(y)=-{\cal G}(-y)$. Therefore, if the outer black hole horizon
coincides with the acceleration horizon then the roots  $x_0$ and
$x_\mathrm{s}$ of ${\cal G}(x)$ also coincide (see Fig.
\ref{g3-flat}). This implies that the range of the angular
coordinate $x$ becomes $x_\mathrm{s}<x\leq x_\mathrm{n}$ since the
proper distance between $x_\mathrm{s}$ and $x_\mathrm{n}$ goes to
infinity \cite{DGKT,ErnstPC}. The point $x_\mathrm{s}$ disappears
from the $x,z$ angular section which is no longer compact but
becomes topologically ${\mathbb{R}}\times S^1$ or
${\mathbb{R}}^2$. So, we have a conical singularity only at
$x=x_\mathrm{n}$ which can be avoided by choosing
$2\kappa^{-1}=|{\cal G}'(x_\mathrm{n})|$. Therefore,  while the
Nariai C-metric ($\Lambda>0$) is topologically conformal to
$dS_2\times \tilde{S}^2$, the ``Nariai"  flat C-metric is
topologically conformal to $dS_2\times {\mathbb{R}}^2$. Its
Carter-Penrose diagram is given by Fig. \ref{nariai-fig}(b). We
could construct the ``Nariai" Ernst solution, however since its
main motivation is related to the removal of the conical
singularities present in the C-metric solution and in this case
our ``Nariai"  flat C-metric is free of conical singularities, we
will not do it.

At this point, it is appropriate to find the $A=0$ limit of the
``Nariai"  flat C-metric. This limit is not direct [see
(\ref{Gfactor L=0})]. To achieve the suitable limit we first make
the coordinate rescales: $\tilde{x}=\chi/A$, $\bar{y}=x/A$, and
$\bar{z}=z/A$. Then, setting $A=0$, the solution becomes
$ds^2=(1+{\cal K})/6\,[{\cal K}^{-1}(-\tilde{x}^2 d\tau^2
+d\tilde{x}^2)+ d\bar{y}^2+ d\bar{z}^2$]. This limit agrees with
the one taken from the $A=0$ Nariai solution (written in
subsection \ref{sec:Nariai}) in the limit $\Lambda=0$ (see
Appendix \ref{sec:lim L=0 Nariai}). Therefore, while the limit
$A=0$ of the Nariai C-metric ($\Lambda>0$) is topologically
$dS_2\times S^2$, the $A=0$ limit of the ``Nariai" flat C-metric
is topologically ${\mathbb{M}}^{1,1}\times {\mathbb{R}}^2$. This
is a reminiscence of the fact that with $\Lambda=0$ when $A$ goes
to zero there is no acceleration horizon to play the role of
cosmological horizon that supports the extremal limit taken in
this subsection. As a final remark in this subsection, we note
that Horowitz and Sheinblatt \cite{HorowShein} have taken a
different extremal limit, which differs from the one discussed in
this subsection mainly because it preserves the asymptotic
behavior  and topology of the original flat C-metric solution.

\subsection{\label{sec:Gen BR L=0}The Bertotti-Robinson  flat C-metric}
Given the Bertotti-Robinson dS C-metric  generated in subsection
\ref{sec:Gen BR dS}, we can construct the Bertotti-Robinson  flat
C-metric by taking the direct $\Lambda=0$ limit of Eqs.
(\ref{br-ds-C-Metric})-(\ref{Gfactor-br}). The gravitational field
of the Bertotti-Robinson flat C-metric is then given by
(\ref{br-ds-C-Metric}) with
\begin{eqnarray}
 &{\cal R}^2(x)=\left (Ax+\sqrt{\frac{6A^2}{1-{\cal K}}}\right
 )^{-2}\:,&
\nonumber \\
  &{\cal G}(x) = 1-x^2
 -\sqrt{\frac{2(1-{\cal K})(2+{\cal K})^2}{27}}\,x^3
 -\frac{1-{\cal K}^2}{12} \,x^4\:,&
 \label{Gfactor-br L=0}
\end{eqnarray}
and $0<{\cal K}< 1$. ${\cal G}(x)$ has only two real roots, the
south pole $x_\mathrm{s}$ and the north pole $x_\mathrm{n}$. The
angular coordinate $x$ can range between these two poles whose
value is calculated in Appendix \ref{sec:angular}. As in the
Bertotti-Robinson dS C-metric, this solution has topology
conformal to  $AdS_2 \times \tilde{S}^2$, and a Carter-Penrose
diagram drawn in Fig. \ref{br-fig}.(b). The solution has a conical
singularity at one of the poles of its deformed 2-sphere
$\tilde{S}^2$. In the $A=0$ limit, we have ${\cal K} \rightarrow
1$ and ${\cal R}^2(x) \rightarrow q^2$ and we obtain the
Bertotti-Robinson solution (\ref{br}) discussed in subsection
\ref{sec:BR}.

From the above solution we can generate the Bertotti-Robinson
Ernst solution whose gravitational field is given by
\begin{eqnarray}
d s^2 &=& \Sigma^2(x)\frac{{\cal R}^2(x)}{{\cal K}}
\left (-\sinh^2\chi\, d\tau^2 +d\chi^2\right ) \nonumber \\
&+&
 {\cal R}^2(x)\left [\frac{\Sigma^2(x)}{{\cal G}(x)}dx^2
 +\frac{{\cal G}(x)}{\Sigma^2(x)}\,dz^2 \right ]
 \:.
 \label{br-Ernst}
\end{eqnarray}
with ${\cal R}^2(x)$ and ${\cal G}(x)$ given by Eq.
(\ref{Gfactor-br L=0}), and with
 \begin{eqnarray}
 \Sigma(x)={\biggl (}1+\frac{1}{2}\, q \, {\cal E}_0\, x{\biggr )}^2
   + \frac{1}{4}{\cal E}_0^2 \, {\cal G}(x){\cal R}^2(x)\:.
 \label{factor-br}
 \end{eqnarray}
 Its electromagnetic field is given by  Eq. (\ref{pot-Ernst-B}), in
 the pure magnetic case, and by Eq. (\ref{pot-Ernst-E}), in
 the pure electric case.
Choosing $\tilde{\kappa}$ satisfying Eq. (\ref{k-s-Ernst}), with
$\tilde{\cal{G}}(x)=\Sigma^{-2}(x){\cal{G}}(x)$,  and ${\cal E}_0$
such that condition (\ref{k=Ernst}) is satisfied, the
Bertotti-Robinson Ernst solution is free of conical singularities.
As a final remark in this subsection, we note that Dowker,
Gauntlett, Kastor and Traschen \cite{DGKT} have taken a different
extremal limit, which differs from the one discussed in this
subsection mainly because it preserves the asymptotic behavior and
topology of the original flat C-metric solution.

\subsection{\label{sec:Gen Nariai-BR L=0}The ``Nariai" Bertotti-Robinson
flat C-metric}

From the Nariai Bertotti-Robinson dS C-metric  generated in
subsection \ref{sec:Gen Nariai-BR dS}, we can construct the
``Nariai" Bertotti-Robinson  flat C-metric by taking the direct
$\Lambda=0$ limit of Eqs.
(\ref{Nariai-C-Metric-N-br})-(\ref{Gfactor-N-br}). We can also
construct the ``Nariai" Bertotti-Robinson Ernst solution. We do
not do these limits here because they are now straightforward.

\section{\label{sec:ExtLim AdS}Extremal limits of
the A\lowercase{d}S C-metric}
In the AdS background there are three different families of
C-metrics. When we set $A=0$, each one of these families reduces
to a different single AdS black hole. Before we deal with the AdS
C-metric we will then briefly describe the main features of each
one of these black holes.

The Einstein equations with $\Lambda<0$ admit a three-family of
black hole solutions whose gravitational field is described by
\begin{eqnarray}
d s^2 = - V(r)\, dt^2 +\frac{dr^2}{V(r)}+r^2 d\Omega_b^2,
 \label{AdS bH}
\end{eqnarray}
with
\begin{eqnarray}
V(R)=b-\frac{2m}{r}+\frac{q^2}{r^2}+\frac{|\Lambda|}{3}r^2 \:,
 \label{N-AdS bh}
\end{eqnarray}
where $b$ can take the values $1,0,-1$ and
\begin{eqnarray} \left\{ \begin{array}{l}
d\Omega_b^2=d \theta^2+\sin^2\theta\,d\phi^2
\:\:\:\:\:\:\:\:{\rm for}\:\: b=1\,,\\
d\Omega_b^2=d \theta^2+d\phi^2
\:\:\:\:\:\:\:\:\:\:\:\:\:\:\:\:\:\:\:\,{\rm for}\:\: b=0 \,,\\
d\Omega_b^2=d \theta^2+\sinh^2\theta\,d\phi^2 \:\:\:\:\:\: {\rm
for}\:\:  b=-1\,.
\end{array} \right.
\label{angular AdS bh}
\end{eqnarray}
These three solutions describe three different kind of black
holes. The black holes with $b=1$ are the usual AdS black holes
with spherical topology. The black holes with $b=0$ have planar,
cylindrical  or toroidal (with genus $g\geq 1$) topology and were
introduced and analyzed in \cite{toroidal}. The topology of the
$b=-1$ black holes is hyperbolic  or, upon compactification,
toroidal with genus $g\geq 2$, and they have been analyzed in
\cite{topological}. These hyperbolic black holes have negative mass. They
have a cosmological horizon (contrarily to what happens with the
$b=1$ and $b=0$ cases) and their charged version has only one
black hole  horizon (as opposed to the charged $b=1$ and $b=0$
cases which have an inner and an outer black hole horizon). The
hyperbolic topology together with the presence of the cosmological
horizon turn these $b=-1$ black holes into the appropriate
solutions that allow the generation of the anti-Nariai solution
(\ref{qNariai-a}) with the limiting Ginsparg-Perry procedure, when
the black hole horizon approaches the cosmological horizon.

Now, to each one of these families corresponds a different AdS
C-metric which has been found by Pleba\'nski and Demia\'nski
\cite{PlebDem}. In what follows we will then describe each one of
these solutions and, in particular, we will generate a new
solution - the anti-Nariai C-metric.

\subsection{\label{sec:ExtLim AdS Sphere}Extremal limits of the
AdS C-metric with spherical horizons}
The gravitational field of the massive charged  AdS C-metric
\cite{PlebDem} is given by Eq. (\ref{C-metric}) with
 \begin{eqnarray}
 & &{\cal F}(y) = \frac{|\Lambda|-3A^2}{3A^2}
                     +y^2-2mAy^3+q^2A^2y^4, \nonumber \\
 & &{\cal G}(x) = 1-x^2-2mAx^3-q^2 A^2 x^4\:,
 \label{FG-spherical}
 \end{eqnarray}
and the electromagnetic field is given by (\ref{F-mag}) and
(\ref{F-el-Lorentz}). This solution describes a pair of
accelerated black holes in the AdS background \cite{OscLem_AdS-C}
when the acceleration $A$ and the cosmological constant are
related by $A>|\Lambda|/3$. When we set $A=0$
\cite{PodEHM,OscLem_AdS-C} we obtain a single non-accelerated AdS
black hole with spherical topology described by Eqs. (\ref{AdS
bH})-(\ref{angular AdS bh}) with $b=1$.

In a way analogous to the one described in last sections we can
generate new solutions from the extremal limits of the AdS
C-metric. Since this follows straightforwardly, we do not do it
here. The new relevant feature of this case is the fact that the
Nariai-like extremal solution only exists when $A>|\Lambda|/3$. In
this case, and only in this one, the acceleration horizon is
present in the AdS C-metric and we can have the outer black hole
horizon approaching it.

\subsection{\label{sec:ExtLim AdS Toroidal}Extremal limits of the
AdS C-metric with toroidal horizons }

The gravitational field of the massive charged toroidal AdS
C-metric (see \cite{PlebDem,MannAdS}) is given by  Eq.
(\ref{C-metric}) with
 \begin{eqnarray}
 & &{\cal F}(y) = \frac{|\Lambda|-3A^2}{3A^2}
                    -2mAy^3+q^2A^2y^4, \nonumber \\
 & &{\cal G}(x) = 1-2mAx^3-q^2 A^2 x^4\:,
 \label{FG-toroidal}
 \end{eqnarray}
and the electromagnetic field is given by (\ref{F-mag}) and
(\ref{F-el-Lorentz}). When we set $A=0$ we obtain the AdS black
hole with planar, cylindrical or toroidal topology described by
Eqs. (\ref{AdS bH})-(\ref{angular AdS bh}) with $b=0$
\cite{toroidal}.

In a way analogous to the one described in section \ref{sec:ExtLim
dS} we can generate new solutions from the extremal limits of the
toroidal AdS C-metric, that are the toroidal AdS counterparts of
the Nariai and Bertotti-Robinson C-metrics. Since this follows
straightforwardly, we do not do it here.

\subsection{\label{sec:ExtLim AdS Topological}Extremal limits of the
AdS C-metric with hyperbolic horizons}

\begin{figure} [th]
\includegraphics*[height=1.6in]{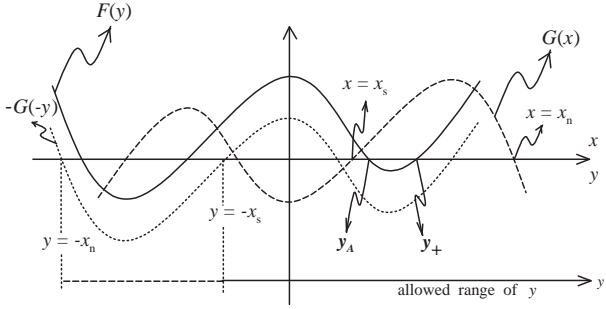}
\caption{\label{g3_topol}
 Shape of ${\cal G}(x)$ and ${\cal F}(y)$ for a general non-extremal
charged massive hyperbolic AdS C-metric studied in section
\ref{sec:ExtLim AdS Topological}. The allowed range of $x$ is
between $x_\mathrm{s}$ and $x_\mathrm{n}$ where ${\cal G}(x)$ is
positive and compact. The permitted range of $y$ is $-x\leq y <
+\infty$. The presence of an accelerated horizon is indicated by
$y_A$ and the black hole horizon by $y_+$. In the anti-Nariai case
considered in subsection \ref{sec:Gen antiNariai}, $y_A$ and $y_+$
coincide. [For completeness we comment on a case not represented
in the figure but discussed on the text: when $q=0$, the zero
$x_\mathrm{n}$ of ${\cal G}(x)$ disappears, and ${\cal G}(x)$
grows monotonically from $x=x_\mathrm{s}$ into $x=+\infty$.]
 }
\end{figure}

The gravitational field of the massive charged AdS
C-metric with hyperbolic horizons, the hyperbolic AdS C-metric,
is given by  Eq. (\ref{C-metric}) with
\cite{PlebDem,MannAdS}
\begin{eqnarray}
& &{\cal F}(y) = \frac{|\Lambda|+3A^2}{3A^2}
                     -y^2-2mAy^3+q^2A^2y^4, \nonumber \\
& &{\cal G}(x) = -1+x^2-2mAx^3-q^2 A^2 x^4\:
\label{FG-topological}
\end{eqnarray}
(represented in Fig. \ref{g3_topol}), and the electromagnetic
field is given by (\ref{F-mag}) and (\ref{F-el-Lorentz}).

This solution depends on four parameters namely, the cosmological
constant $\Lambda<0$, the acceleration parameter $A>0$, and $m$
and $q$ which are mass and electromagnetic charge parameters,
respectively. When $A=0$ this solution reduces to the hyperbolic
black holes [case $b=-1$ discussed in Eqs. (\ref{AdS
bH})-(\ref{angular AdS bh})].

\subsubsection{\label{sec:Gen antiNariai}The anti-Nariai C-metric}

We are interested in a particular extreme  hyperbolic AdS
C-metric, for which $y_A=y_+$ (see Fig. \ref{g3_topol}), and let
us label this degenerated horizon by $\rho$: $y_A=y_+\equiv \rho$.
In this case, the function ${\cal F}(y)$ can be written as
\begin{eqnarray}
{\cal F}(y) =-\frac{\rho^2-3\gamma}{\rho^4}
 (y-y_{\rm neg})(y-y'_{\rm neg})(y-\rho)^2\:,
 \label{Fextreme-a}
 \end{eqnarray}
where
\begin{eqnarray}
\gamma =\frac{|\Lambda|+3A^2}{3A^2}\:,
 \label{gamma-a}
 \end{eqnarray}
and the degenerate root $\rho$, and the negative roots $y_{\rm
neg}$ and $y'_{\rm neg}$ are given by
\begin{eqnarray}
& & \rho =\frac{3m}{4q^2A}
 \left ( 1+ \sqrt{1+\frac{8}{9}\frac{q^2}{m^2}} \:\right )
 \:,  \nonumber \\
& & y_{\rm neg} =-\frac{\gamma \rho}{3\gamma-\rho^2}
 \left ( 1+ \sqrt{\frac{\rho^2-2\gamma}{\gamma}} \:\right )
 \:, \nonumber \\
& & y'_{\rm neg} =-\frac{\gamma \rho}{3\gamma-\rho^2}
 \left ( 1- \sqrt{\frac{\rho^2-2\gamma}{\gamma}} \:\right )
 \:.
 \label{zerosy3-a}
 \end{eqnarray}
The mass parameter and the charge parameter of the solution are
written as a function of $\rho$ as
\begin{eqnarray}
& &m =-\frac{1}{A\rho}
 \left ( 1- \frac{2\gamma}{\rho^2} \right )
 \:,  \nonumber \\
& & q^2 =\frac{1}{A^2\rho^2}
 \left ( \frac{3\gamma}{\rho^2}-1 \right )
 \:.
 \label{mq-a}
 \end{eqnarray}
The requirement that $y_{\rm neg}$ and $y'_{\rm neg}$ are real roots
and the condition $q^2 \geq 0$ require that the allowed range of
$\rho$ is
\begin{eqnarray}
 2\gamma < \rho^2 \leq 3\gamma \:.
 \label{range-gamma-a}
\end{eqnarray}
The mass and the charge of the anti-Nariai type solution,
$m_{\rm aN}$ and $q_{\rm aN}$, respectively,
are both
monotonically decreasing functions of $\rho$, and as one comes from
$\rho=\sqrt{3\gamma}$ into $\rho=\sqrt{2\gamma}$ one has,
\begin{eqnarray}
& &  -\frac{1}{3}
  \frac{1}{\sqrt{|\Lambda|+3A^2}}\leq m_{\rm aN}<0\:, \nonumber \\
& & 0\leq q_{\rm aN}< \frac{\sqrt{3}}{2}
  \frac{1}{\sqrt{|\Lambda|+3A^2}}\:.
 \label{mq-antiNariai}
 \end{eqnarray}

In order to generate the anti-Nariai C-metric from the
near-extreme topological AdS C-metric we first set
\begin{eqnarray}
 y_A=\rho-\varepsilon, \:\:\:\:y_+=\rho+\varepsilon,
 \:\:\:\:\:\:{\rm with} \:\: \varepsilon<<1\:,
 \label{NariaiLimit-a}
\end{eqnarray}
in order that $\varepsilon$ measures the deviation from
degeneracy, and the limit $y_A\rightarrow y_+$ is obtained when
$\varepsilon \rightarrow 0$. Now, we introduce a new time
coordinate $\tau$ and a new radial coordinate $\chi$,
 \begin{eqnarray}
t= \frac{1}{\varepsilon {\cal K}}\,\tau \:, \:\:\:\:\:\:\:\:\:\:\:
y=\rho+\varepsilon \cosh\chi \:,
 \label{NariaiCoord-a}
\end{eqnarray}
where
\begin{eqnarray}
& & \!\!\! {\cal K} = \frac{3\gamma-\rho^2}{\rho^4}
 (\rho-y_{\rm neg})(\rho-y'_{\rm neg})=
 \frac{2(|\Lambda|+3A^2)}{A^2\rho^2}-1\:,  \nonumber \\
& &
 \label{Kfactor-a}
\end{eqnarray}
and $2\gamma < \rho^2 \leq 3\gamma$ implies $1 \leq {\cal K}<2$
with $q=0\Rightarrow {\cal K}=1$. In the limit $\varepsilon
\rightarrow 0$, from Eq. (\ref{C-metric}) and Eq.
(\ref{Fextreme-a}), the metric becomes
\begin{eqnarray}
d s^2 &=& \frac{{\cal R}^2(x)}{{\cal K}}
\left (-\sinh^2\chi\, d\tau^2 +d\chi^2\right ) \nonumber \\
& & +
 {\cal R}^2(x)\left [{\cal G}^{-1}(x)dx^2+ {\cal G}(x)dz^2 \right ]
 \:.
 \label{Nariai-C-Metric-a}
\end{eqnarray}
where
\begin{eqnarray}
 &{\cal R}^2(x)= \left (
Ax+\sqrt{\frac{2(|\Lambda|+3A^2)}{1+{\cal K}}}\right )^{-2}\:,& \\
\label{Rfactor-a}
  &  \!\!\!{\cal G}(x) = -1+x^2
+\frac{A}{3} \sqrt{\frac{2(1+{\cal K})(2-{\cal
K})^2}{(|\Lambda|+3A^2)}}\,x^3
  +\frac{A^2}{4}\frac{{\cal K}^2-1}{|\Lambda|+3A^2} \,x^4\:,&
  \nonumber \\
 \label{Gfactor-a}
\end{eqnarray}
Under the coordinate transformation (\ref{NariaiCoord-a}), the
Maxwell field for the magnetic case is still given by Eq.
(\ref{F-mag}), while in the electric case, (\ref{F-el-Lorentz})
becomes
 \begin{eqnarray}
 F=-\frac{q}{{\cal K}}\,\sinh \chi \,d\tau\wedge d\chi\:.
\label{F-el-antiNariai}
\end{eqnarray}
The Carter-Penrose diagram of the anti-Nariai C-metric is also
given by Fig. \ref{br-fig}.(b).

At this point, let us focus our attention on the angular surfaces
with $\tau=$constant and $\chi=$constant. When $q\neq 0$, we
choose $x$ such that it belongs to the range
$[x_\mathrm{s},x_\mathrm{n}]$ (sketched in Fig. \ref{g3_topol})
where ${\cal G}(x)\geq 0$. In this way, the metric has the correct
signature, the angular surfaces are compact, and the allowed range
of $y$ includes the acceleration ($y_A$) and black hole ($y_+$)
horizons (if we had chosen the other possible interval of $x$
where ${\cal G}(x)\geq 0$, sketched in Fig. \ref{g3_topol}, this
last condition would not be satisfied). We unavoidably have a
conical singularity at least at one of the poles of this compact
surface (that we label by $\tilde{S}^2$, say). Thus, the charged
anti-Nariai C-metric has a compact angular surface of fixed size
with a conical singularity at one of its poles. It is
topologically conformal to $AdS_2\times \tilde{S}^2$. When $q=0$,
the zero $x_\mathrm{n}$ of ${\cal G}(x)$ disappears, and ${\cal
G}(x)$ grows monotonically from $x=x_\mathrm{s}$ into $x=+\infty$.
Then, the angular surfaces are not compact, and we have a single
pole ($x=x_\mathrm{s}$) with a conical singularity, which can be
eliminated. Thus, the neutral anti-Nariai C-metric has a
non-compact angular surface of fixed size (a kind of a deformed
2-hyperboloid that we label by $\tilde{H}_2$, say) which is free
of conical singularities. It is topologically conformal to
$AdS_2\times \tilde{H}_2$.

In order to obtain the $A=0$ limit, we first set
$\hat{\rho}=A\rho$ [see first relation of (\ref{zerosy3-a})], a
parameter that has a finite and well-defined value when
$A\rightarrow 0$. Then when $A\rightarrow 0$ we have ${\cal K}
\rightarrow {\cal K}_0=2|\Lambda|/\hat{\rho}^2-1$ and ${\cal R}^2
\rightarrow {\cal R}_0^{\:2}=\hat{\rho}^{-2}$, with ${\cal
R}_0^{\:2}$ and ${\cal K}_0$ satisfying relations
(\ref{relations-a}). Moreover, when we set $A = 0$, $m\neq 0$ and
$q\neq 0$, the coordinate transformations $\theta =
\int_{x_\mathrm{s}}^{x}{\cal{G}}^{-1/2}dx$ and $\phi=z$ imply that
$x\in[x_\mathrm{s}=+1,+\infty[$, $x=\cosh \theta$, and ${\cal
G}=-1+x^2=\sinh^2 \theta$. The angular surface then reduces to a
2-hyperboloid, $H_2$, of fixed size with line element
$d\theta^2+\sinh^2{\theta}\,d\phi^2$. Therefore, when $A=0$ the
anti-Nariai C-metric reduces to the anti-Nariai solution
(\ref{qNariai-a}) described in subsection \ref{sec:anti-Nariai},
with topology $AdS_2 \times H_2$. The limiting procedure that has
been applied in this subsection has generated a new exact solution
that satisfies the Maxwell-Einstein equations in a negative
cosmological constant background.

\subsubsection{\label{sec:other limits} Other extremal limits}

We could also
discuss other extremal limits of the charged topological AdS
C-metric (see Fig. \ref{g3_topol}), but these do not seem to be so
interesting.

\section{\label{sec:Conc}Conclusions}

Following the limiting approach first introduced by Ginsparg and
Perry \cite{GinsPerry}, we have analyzed the extremal limits of
the dS, flat and AdS C-metrics. Among other new
exact solutions, we have generated the Nariai C-metric, the
Bertotti-Robinson C-metric, the Nariai Bertotti-Robinson C-metric
and the anti-Nariai C-metric. These solutions are the C-metric
counterparts of the well know solutions found in the 1950's. They
are specified by an extra parameter: the acceleration parameter
$A$ of the C-metric from which they are generated. When we set
$A=0$ the solutions found in this paper reduce to the Nariai, the
Bertotti-Robinson, the Nariai Bertotti-Robinson and the
anti-Nariai solutions.

One of the features of these $A=0$ solutions is the fact that they
are topologically the direct product of two 2-dimensional
manifolds of constant curvature. Their C-metric counterparts are
conformal to this topology, with the conformal factor depending on
the angular coordinate. Moreover, the angular surfaces of these
new C-solutions have a fixed size, but they lose the symmetry of
the $A=0$ counterparts. For example, while the angular surfaces of
the Nariai and Bertotti-Robinson solutions are round 2-spheres,
the angular surfaces of their C-metric counterparts are deformed
2-spheres - they are compact but not round. Another important
difference between the $A=0$ and $A \neq 0$ solutions is the fact
that the $A \neq 0$ solutions have, in general, a conical
singularity at least at one of the poles of their angular
surfaces. This conical singularity is a reminiscence of the
conical singularity that is present in the C-metric from which
they were generated. In the C-metric these conical singularities
are associated to the presence of a strut or string that furnishes
the acceleration of the near-extremal black holes. In this
context, we find that the Nariai C-metric generated from a
extremal limit of the dS C-metric describes  a spacetime that is
conformal to the product $dS_2\times \tilde{S}^2$. To each point
in the deformed 2-sphere corresponds a $dS_2$ spacetime, except
for one point which corresponds a $dS_2$ spacetime with an
infinite straight strut or string, with a mass density and
pressure satisfying $p=-\mu$. Analogously, the Nariai
Bertotti-Robinson dS C-metric describes a spacetime that is
conformal to the product ${\mathbb{M}}^{1,1} \times \tilde{S}^2$.
To each point in the deformed 2-sphere corresponds a
${\mathbb{M}}^{1,1}$ spacetime, except for one point which
corresponds a ${\mathbb{M}}^{1,1}$ spacetime with an infinite
straight strut or string. In the case of the Bertotti-Robinson dS
C-metric (topologically conformal to $AdS_2\times \tilde{S}^2$),
the strut or string does not survive to the Ginsparg-Perry
limiting procedure, and thus in the end of the process we only
have a conical singularity.

In what concerns the causal structure, the Carter-Penrose diagrams
of the $A \neq 0$ solutions are equal to those of the $A=0$
solutions. For example, the diagram of the Nariai C-metric is
equal to the one that describes the (1+1)-dimensional dS
solution, the diagram of the Bertotti-Robinson C-metric and of the
anti-Nariai C-metric is equal to the one that describes the
(1+1)-dimensional AdS solution, and the diagram of the
Nariai Bertotti-Robinson C-metric is given by the Rindler diagram.

Some of these solutions, perhaps all, are certainly of physical
interest.  Indeed, it is known that the Nariai solution ($A=0$) is
unstable and, once created, it decays through the quantum
tunnelling process into a slightly non-extreme black hole pair
\cite{Bousso60y}. We then expect that the Nariai C-metric is also
unstable and that it will decay into a slightly non-extreme pair
of black holes accelerated by a strut or by a string. The
solutions found in this paper also play an important role in the
decay of the dS or AdS spaces, and therefore can mediate the
Schwinger-like quantum process of pair creation of black holes.
Indeed, the Nariai, and the dS Nariai Bertotti-Robinson instantons
($A=0$) are one of the few Euclidean solutions that are regular,
and have thus been used
\cite{MelMosRom,MannRoss,BooMann,VolkovWipf} to study the pair
creation of dS black holes materialized and accelerated by the
cosmological constant background field (an instanton is a solution
of the Euclidean field equations that smoothly connects the
spacelike sections of the initial state, the pure dS space in this
case, and the final state, the dS black hole pair in this case).
Moreover, the Euclidean ``Nariai" flat C-metric \cite{flatPC} and
Ernst solution \cite{DGKT,ErnstPC} (also discussed in this paper)
have been used to analyze the process of pair production of
$\Lambda=0$ black holes, accelerated by a string or by an
electromagnetic external field, respectively. Therefore, its
natural to expect that the Euclidean extremal limits of the dS
C-metric and AdS C-metric found in this paper mediate the process
of pair creation of black holes in a cosmological background, that
are then accelerated by a string, in addition to the cosmological
field acceleration. For the Nariai case, e.g., the picture would
be that of the nucleation, in a dS background, of a Nariai C
universe, whose string then breaks down and a pair of dS black
holes is created at the endpoints of the string. This expectation
is confirmed in \cite{OscLem-PCdS}.


\begin{acknowledgments}

This work was partially funded by Funda\c c\~ao para a Ci\^encia e
Tecnologia (FCT) through project CERN/FIS/43797/2001 and
PESO/PRO/2000/4014.  OJCD also acknowledges finantial support from the
FCT through PRAXIS XXI programme. JPSL acknowledges finantial support
from ICCTI/FCT and thanks Observat\'orio Nacional do Rio de Janeiro
for hospitality.

\end{acknowledgments}

\appendix
\section{\label{sec:lim L=0 Nariai}\bm{$\Lambda=0$}
limit of the Nariai and anti-Nariai solutions}
In this appendix we find the $\Lambda=0$ limit of the Nariai
solution, Eq. (\ref{qNariai}),  and of the anti-Nariai solution,
Eq. (\ref{qNariai-a}). In this limit the line element of both
solutions goes apparently to infinity since ${\cal
R}_0^{\:2}\rightarrow \infty$. To achieve the suitable limit of
the Nariai solution, we first make the coordinate rescales:
$\tilde{x}= ({\cal R}_0/ \sqrt{{\cal K}_0})\chi$, and
$\varrho={\cal R}_0 \,\theta$. Then, taking the $\Lambda=0$ limit,
the solution becomes
\begin{eqnarray}
d s^2 = (-\tilde{x}^2 d\tau^2 +d\tilde{x}^2)+ (d\varrho^2+
\varrho^2\,d\phi^2) \, .
 \label{Lim L=0 Nariai1}
\end{eqnarray}
The spacetime sector is a Rindler factor, and the angular sector
is a cylinder. Therefore, under the usual coordinate
transformation $\tilde{x}=\sqrt{\bar{x}^2-\bar{t}^2}$ and
$\tau={\rm arctanh(\bar{t}/\bar{x})}$, and unwinding the cylinder
($\bar{y}=\varrho\,\cos\phi $ and $\bar{z}=\varrho\,\sin\phi$), we
have
\begin{eqnarray}
d s^2 = (-d\bar{t}^{\,2} +d\bar{x}^2)+ (d\bar{y}^2+ d\bar{z}^2)\,
.
 \label{Lim L=0 Nariai2}
\end{eqnarray}
Therefore, while the Nariai C-metric is topologically $dS_2\times
S^2$, its $\Lambda=0$ limit  is topologically
${\mathbb{M}}^{1,1}\times {\mathbb{R}}^2$, i.e.,
${\mathbb{R}}^4$.

A similar procedure shows that taking the $\Lambda=0$ limit of the
anti-Nariai solution (\ref{qNariai-a}) ($AdS_2\times H_2$) leads
to Eq. (\ref{Lim L=0 Nariai2}).

\section{\label{sec:angular}Determination of the north and south poles}

In this appendix we discuss the zeros of the function ${\cal
G}(x)$ that appears in the extremal limits of the dS C-metric, and
in the extremal limits of the flat C-metric. This function ${\cal
G}(x)$ has only two real zeros in the cases discussed in this
paper, namely the Nariai, the Bertotti-Robinson, and the Nariai
Bertotti-Robinson (both for $\Lambda>0$ and $\Lambda=0$). These
two roots are the south pole and the north pole, and are
respectively given by
\begin{eqnarray}
x_\mathrm{s} & = & -p + \frac{h}{2} - \frac{a}{4\,b} < 0\, , \nonumber \\
x_\mathrm{n} & = & +p +
      \frac{h}{2} - \frac{a}{4\,b} > 0 \:,
\label{polos - cold}
    \end{eqnarray}
with
\begin{eqnarray}
p & = &\frac{1}{2}\left (-\frac{s}{3} + \frac{a^2}{2\,b^2} -
 \frac{1 - 12 \,b}{3\,s\,b^2} - \frac{4}{3\, b} + n\right)^{1/
          2} \:, \nonumber \\
n & = &\frac{-a^3 + 4\,a\, b}{4\,h \,b^3} \:, \nonumber \\
h & = &\sqrt{\frac{s}{3} + \frac{a^2}{4 \,b^2} +
   \frac{1 -12\, b}{3\,s\, b^2} - \frac{2}{3\,b}} \:, \nonumber \\
s & = &\frac{1}{2^{1/3}\, b}\left ( \lambda - \sqrt{\lambda^2 -
 4(1-12\, b)^3} \right)^{1/
          3} \:, \nonumber \\
        \lambda & = & 2 - 27 \,a^2 + 72\, b\:,
\label{acess - zeros - ang}
     \end{eqnarray}
where $a$ and $b$ are, respectively, the absolute values of the
coefficients of $x^3$ and $x^4$ in Eqs. (\ref{Gfactor}),
(\ref{Gfactor-br}), (\ref{Gfactor-N-br}), (\ref{Gfactor L=0}), and
(\ref{Gfactor-br L=0}). For the function ${\cal G}(x)$ written in
a different polynomial form that facilitates the determination of
its zeros see Hong and Teo \cite{HongTeo}.


\end{document}